\documentclass[twocolumn]{aastex631}
\usepackage{graphicx}

\usepackage{subfigure}
\usepackage{soul}

\newcolumntype{H}{>{\setbox0=\hbox\bgroup}c<{\egroup}@{}}

\shorttitle{VLA FRAMEx~I: NGC~4388}
\shortauthors{Sargent et al.}

\defcitealias{2021ApJ...906...88F}{FRAMEx~I}
\defcitealias{2022ApJ...927...18F}{FRAMEx~II}
\defcitealias{2022ApJ...936...76S}{FRAMEx~III}
\defcitealias{2023ApJ...958...61F}{FRAMEx~IV}

\begin{document}

\title{VLA FRAMEx. I. Wideband Radio Properties of the AGN in NGC~4388}

\email{andrew.j.sargent2.civ@us.navy.mil}
\AuthorCollaborationLimit=8

\author[0000-0002-8049-0905]{Andrew J.\ Sargent}
\affiliation{United States Naval Observatory, 3450 Massachusetts Ave., NW, Washington, DC 20392, USA}
\affiliation{Department of Physics, The George Washington University, 725 21st St. NW, Washington, DC 20052, USA}

\author[0000-0002-3365-8875]{Travis C.\ Fischer}
\affiliation{AURA for ESA, Space Telescope Science Institute, 3700 San Martin Drive, Baltimore, MD 21218, USA}

\author[0000-0002-4146-1618]{Megan C. Johnson}
\affiliation{United States Naval Observatory, 3450 Massachusetts Ave., NW, Washington, DC 20392, USA}

\author[0000-0001-9149-6707]{Alexander J.\ van der Horst}
\affiliation{Department of Physics, The George Washington University, 725 21st St. NW, Washington, DC 20052, USA}

\author[0000-0002-4902-8077]{Nathan J.\ Secrest}
\affiliation{United States Naval Observatory, 3450 Massachusetts Ave., NW, Washington, DC 20392, USA}

\author[0000-0003-4727-2209]{Onic I. Shuvo}
\affiliation{Department of Physics, University of Maryland Baltimore County, 1000 Hilltop Circle, Baltimore, MD 21250, USA}

\author[0000-0002-8736-2463]{Phil J.\ Cigan}
\affiliation{United States Naval Observatory, 3450 Massachusetts Ave., NW, Washington, DC 20392, USA}

\author[0000-0001-5785-7038]{Krista L.\ Smith}
\affiliation{Department of Physics and Astronomy, Texas A\&M University, College Station, TX 77843, USA}

\begin{abstract}

We present the first results from Karl G. Jansky Very Large Array (VLA) observations as a part of the Fundamental Reference Active Galactic Nucleus (AGN) Monitoring Experiment (FRAMEx), a program to understand the relationship between AGN accretion physics and wavelength-dependent position as a function of time.
With this VLA survey, we investigate the radio properties from a volume-complete sample of 25 hard X-ray-selected AGNs using the VLA in its wideband mode. We observed the targets in the A-array configuration at 4--12 GHz with all polarization products. In this work, we introduce our calibration and imaging methods for this survey, and we present our results and analysis for the radio quiet AGN NGC~4388. We calibrated and imaged these data using the multi-term, multi-frequency synthesis imaging algorithm to determine its spatial, spectral and polarization structure across a continuous $4-12$ GHz band. In the AGN, we measure a broken power law spectrum with $\alpha=-0.06$ below a break frequency of 7.3 GHz and $\alpha=-0.34$ above. We detect polarization at sub-arcsecond resolution across both the AGN and a secondary radio knot. We compare our results to ancillary data and find that the VLA radio continuum is likely due to AGN winds interacting with the local interstellar medium that gets resolved away at sub-parsec spatial scales as probed by the Very Long Baseline Array. A well-known ionization cone to the southwest of the AGN appears likely to be projected material onto the underside of the disk of the host galaxy.

\end{abstract}

\section{Introduction} \label{sec:intro}

Active Galactic Nuclei (AGNs) are morphologically complex objects that can vary dramatically in appearance as a function of orientation with respect to the line of sight, wavelength, and time. In general, AGNs are thought to have a supermassive black hole (SMBH) surrounded by a thermal accretion disk that peaks in the ultraviolet (UV) on scales of a few light-days from the event horizon, the last stable orbit around the SMBH \citep[e.g.,][]{2022ApJ...929...19G}, a broad line region (BLR) of high-velocity gas on scales of tens to hundreds of light-days \citep[e.g.,][]{2013ApJ...767..149B, 2022ApJ...929...19G}, and dense molecular gas orbiting at pc scales \citep[e.g.,][]{2019A&A...623A..79C} that is likely responsible for AGN reddening and obscuration. On tens to thousands of pc scales, ionizing radiation from the accretion disk produces strong forbidden line emission such as [\ion{O}{3}] in the host galaxy's interstellar medium (ISM), producing the ``narrow line region'' (NLR) \citep[e.g.,][]{2003ApJ...597..768S,2006AJ....132..546G}. The NLR also typically shows synchrotron emission at GHz frequencies due to jets, shocks from AGN disk winds, or nearby star formation, and the kinematics and power of the NLR is strongly correlated with GHz radio luminosity \citep[e.g.,][]{2013MNRAS.433..622M}. Time-domain reverberation studies have shown that the closest physical structure to the SMBH is the hot X-ray corona, a compact structure of uncertain morphology that exists a few gravitational radii from the event horizon and is responsible for inverse-Compton scattering of disk UV photons to $\sim1-100$~keV energies \citep[e.g.,][and references therein]{2015MNRAS.451.4375F}. The X-ray corona may also be the origin of core radio emission in ``radio-quiet'' AGNs, which comprise $\sim90\%$ of all AGNs \citep[for a recent review, see][]{2019NatAs...3..387P}. The radio emission from $\sim10\%$ of AGNs that are ``radio-loud''  (e.g., the ICRF radio sources) is likely powered by a relativistic jet, the production of which is a topic of ongoing study, but is likely connected to SMBH spin \citep[as reviewed in][]{2019ARA&A..57..467B}. 
High-resolution X-ray studies have demonstrated that the intergalactic medium between galaxies in dense clusters is likely to be heated significantly by relativistic jets, underscoring the role of AGN ``feedback'' in galaxy evolution \citep[e.g.,][]{2012ARA&A..50..455F}.

Studies of nearby galaxies often allow for the location and morphologies of the various accretion and emission structures seen in AGNs to be resolved. This is not the case for moderate-redshift ($z\sim1$) AGNs, where $1\arcsec$ corresponds to ${<}10$~kpc, in which only the relativistic jets are resolvable. The unresolved structure on pc scales shows evidence of wavelength and time-dependent apparent positions, which creates a problem for efforts to produce a multi-wavelength celestial reference frame.
Consequently, the enabling of precision comparable to radio Very Long Baseline Interferometry (VLBI) in optical astrometry by Gaia led to confirmation that $\sim15\%$ of the radio AGNs that comprise the International Celestial Reference Frame (ICRF) have significantly offset optical and radio positions, typically differing by $0.1-1$~mas ($\sim0.8-8$~pc at $z=1$).
While optical and radio position angles may be correlated in VLBI sources with measurable jet angles \citep{2017A&A...598L...1K}, the nature of these offsets is likely complex and requires a more complete picture of AGN accretion and the relationship between AGNs and their host galaxies.

Towards this goal, the Fundamental Reference AGN Monitoring Experiment (FRAMEx) is an observational campaign led by the U.S.\ Naval Observatory to study the apparent radio positions and morphologies of AGNs by probing AGN cores at sub-parsec scales.
The FRAMEx campaign utilized the 105-month Neil Gehrels Swift Observatory (Swift) Burst Alert Telescope (BAT) Catalog \citep{2018ApJS..235....4O} to construct a volume-complete sample of 25 nearby ($<40$~Mpc) AGNs with hard X-ray (14–195 keV) luminosities above $10^{42} {\rm ~erg~s^{-1}}$. This hard X-ray luminosity threshold corresponds to a radio luminosity of $10^{36}~\rm{erg~s^{-1}}$ for a $10^6$ solar mass BH according to the Fundamental Plane of black hole activity \citep[FP;][]{2003MNRAS.345.1057M}.
\cite{2021ApJ...906...88F} (hereafter \citetalias{2021ApJ...906...88F}) campaign obtained Very Long Baseline Array (VLBA) observations at 6~GHz contemporaneously with the Swift X-Ray Telescope (XRT) for the 25 targets and the results were surprising: the VLBA detected only 9 out of the 25 target sources despite deep integrations to achieve root mean square (rms) noise levels of $\sim20$~$\mu$Jy. The corresponding radio luminosity upper limits for the 16 non-detections were far below the FP predictions, despite the same objects appearing to follow the FP when using archival Karl G. Jansky Very Large Array (VLA) data at lower resolution (see Figure 4 of \citetalias{2021ApJ...906...88F}).

Since \citetalias{2021ApJ...906...88F}, several follow-up observing campaigns have been carried out. FRAMEx performed a monitoring campaign for the 9 VLBA detections, observing every 28 days for six epochs, again with quasi-simultaneous Swift XRT and VLBA observations. The first results on the X-ray-variable Seyfert NGC~2992 showed anti-correlated variability between the core radio and $2-10$~keV X-ray emission \citep[][hereafter \citetalias{2022ApJ...927...18F}]{2022ApJ...927...18F}, suggesting an increase in free-free absorption in the BLR co-temporal with an X-ray flare, possibly due to an accretion disk outburst.

Another follow-up expanded the FRAMEx sample with snapshot observations of 9 new objects to include targets with both cosmological and redshift-independent distances less than 40~Mpc, resulting in a total of 34 objects, and reobserved 9 of the 16 objects not detected by the VLBA in \citetalias{2021ApJ...906...88F} \citep[][hereafter \citetalias{2022ApJ...936...76S}]{2022ApJ...936...76S}. The reobservations were again done with the VLBA, but with the integration time quadrupled in order to double the sensitivity as compared to the original observations. Despite very deep observations, rms noise levels of $10$~$\mu$Jy,
only 3 out of the 9 targets were detected. The strict upper limits on their $L_R/L_X$ ratios are as low as $<10^{-8}$ for these hard X-ray-selected AGNs, and \citetalias{2022ApJ...936...76S} analysis suggests that these ``radio silent'' objects may be experiencing strong synchrotron self-absorption (if the object has a high brightness temperature) or free-free absorption in their cores as observed in NGC~2992 in \citetalias{2022ApJ...927...18F}.

The lack of detections for many of the FRAMEx targets at the spatial scales of the VLBA versus the archival data for those same targets using the VLA is peculiar, especially in light of results that the same population of AGNs has a nearly $\sim100\%$ detection rate at 22~GHz with the VLA \citep{2020MNRAS.492.4216S}. If the FP scaling relationship is indeed universally true for accreting black holes, then we might expect it to be invariant to changes in angular resolution. On the other hand, the \citetalias{2021ApJ...906...88F} result implies that any relationship between the observed radio and X-ray luminosities in AGNs develops on larger scales in emission resolved out by the VLBA, calling into question a direct causal relationship with black hole mass, at least for the SMBHs that power AGNs.

The archival VLA observations discussed in \citetalias{2021ApJ...906...88F}, however, come from a time range spanning 30+ years, all with differing flux limits, array configurations and science goals. In 2012, the VLA was upgraded to the Karl G. Jansky VLA and now has a 100\% frequency coverage from 1-50 GHz and can simultaneously measure up to 4 GHz in bandwidth in its dual polarization mode. Therefore, the VLA was chosen to observe the 25 \citetalias{2021ApJ...906...88F} targets at its most extended A-array configuration with full-Stokes polarization across $4-12$ GHz. These wide bandwidth observations provide an opportunity to fill in the large angular disparity from the archival VLA data and simultaneously obtain a uniform data set across all 25 FRAMEx targets. With the A-array configuration, we probe the highest spatial resolution of the VLA ($\sim0\farcs3$ at 5 GHz) such that we can compare to the largest angular scale of the VLBA ($\sim0\farcs05$ at 5 GHz). Simultaneously, these radio continuum observations will allow us to more stringently determine the emission processes and magnetic field properties and at what physical scales they occur. The wideband VLA observations provide the sensitivity and $uv$-coverage to study the impact of the AGN on the nuclear environment on spatial scales ranging from tens to thousands of pc.

In this work, we present the first results from this new observational campaign for the target NGC~4388 and analyze its spatial, spectral, and polarization features. Table \ref{tab:target_parameters} shows parameters for NGC 4388 used in our analysis. The observations and data calibration for NGC~4388 are discussed in Section~\ref{sec:methodology}, we present our results in Section~\ref{sec:results}, and provide analysis of these results in comparison to ancillary data in Section~\ref{sec:discussion}.

\begin{deluxetable}{rcc}
    \tablecaption{NGC~4388}
    \tablehead{Parameter & Value & Source}
    \centering
    \startdata
    Type                           &           Sy2                             & 1\\
    Redshift                      &           0.0084                           & 1\\
    Distance           &           $16.8~{\rm Mpc}$                & 2\\
    $\log(M_{{\rm  BH}})$          &           $6.94~M_{\odot}$                & 1\\
    $\log(L_{\rm{X,~2-10~keV}})$   &           $43.03~{\rm erg~s^{-1}}$        & 1\\
    $L_{{\rm   H\alpha}}$    &           $(0.39\pm0.06)\times10^{41}~{\rm erg~s^{-1}}$ & 3\\
    ${\rm SFR}$              &           $2.42\pm0.23~M_{\odot}~{\rm yr}^{-1}$   & 3\\
    ${\rm SNR_{max}}$              &           $5.9\times10^{-2}~{\rm yr}^{-1}$ & 1
    \enddata
    \tablecomments{NGC~4388 primary target parameters.\\
    (1) \cite{2021ApJ...906...88F}, Table 1. \\
    (2) \cite{2003MNRAS.345.1057M}, Table 1.\\
    (3) \cite{2019ApJ...881...26V}, Table 2.}
    \label{tab:target_parameters}
\end{deluxetable}

\section{Methodology} \label{sec:methodology}
\subsection{Observations} \label{sec:observations}
Our VLA FRAMEx data was observed under the project code 20B-241 (P.I. Fischer). $X$-band ($8-12~\rm{GHz}$, scheduling block 39382198) observations of NGC~4388 occurred on 2 March 2021 and $C$-band ($4-8~\rm{GHz}$, scheduling block 39381834) observations on 3 March 2021. Table \ref{tab:observations} indicates the calibration and target sky positions and on-source times and Table \ref{tab:parameters} shows the observing parameters. First, the primary calibrator 3C~286 was observed for absolute flux density calibrations and polarization angle calibration. This was followed by an observation of the unpolarized source J1310+3220 to calibrate for polarization leakage between the cross-hands. We then used phase referencing with the phase calibrator source J1254+1141 near the target and first observed for 6 minutes and then alternated 9 minute observations of the target source with 2-minute observations of the phase calibrator. On-source time totaled 45 minutes per band. This provides for a theoretical noise level 
of ${\sim}17~{\rm \mu Jy~beam^{-1}}$ per 128 MHz spectral window for 25 antennas assuming a system equivalent flux distribution (SEFD) of 300 Jy, a correlator efficiency of 0.93 and a 15\% data loss due to Radio Frequency Interference (RFI). The VLBA snapshot observations in \citetalias{2021ApJ...906...88F} targeted a $5\sigma$ detection of ${\sim}100~{\rm\mu Jy~beam^{-1}}$ across a 512 MHz band.

\begin{deluxetable*}{rcCCcc}
    \tablecaption{Observation Targets}
    \centering
    \tablehead{ & \colhead{Target}  & \colhead{R.A.} & \colhead{Decl.}  & \colhead{$T_{{\rm int},C}$}   & \colhead{$T_{{\rm int},X}$}}
    \startdata
    Primary Target                     & NGC~4388      & 12^{\rm h}25^{\rm m}46\fs 810000 & +12\degr39\arcmin43\farcs 40000 & 2612 s & 2616 s \\
    Primary Flux Calibrator            & 3C 286        & 13^{\rm h}31^{\rm m}08\fs 287984 & +30\degr30\arcmin32\farcs 95886 & ~974 s & ~972 s \\
    Polarization Angle Calibrator      & 3C 286        & 13^{\rm h}31^{\rm m}08\fs 287984 & +30\degr30\arcmin32\farcs 95886 & ~974 s & ~972 s \\
    Polarization Leakage Calibrator    & J1310+3220    & 13^{\rm h}10^{\rm m}28\fs 663845 & +32\degr20\arcmin43\farcs 78294 & ~774 s & ~774 s \\
    Primary Phase Calibrator           & J1254+1141    & 12^{\rm h}54^{\rm m}38\fs 255601 & +11\degr41\arcmin05\farcs 89507 & ~964 s & ~960 s
    \enddata
    \tablecomments{The calibration and primary targets and their positions are described in Columns $1-4$. The $C$- and $X$-band integration times are in Columns  5 and 6, respectively.}
    \label{tab:observations}
\end{deluxetable*}

\begin{deluxetable}{rc}    \label{tab:parameters}
    \tablecaption{Observation Parameters}
    \tablehead{Parameter & Value}
    \centering
    \startdata
    VLA Configuration & A$\rightarrow$D\\
    Longest Baseline & 36.6 km\\
    Shortest Baseline ($C$-, $X$-band) & 40.0 m, 129.0 m\\
    Number of antennas ($C$-, $X$-band) & 26, 25\\
    Total bandwidth per band & 4 GHz \\
    Number of IF windows per band & 32 \\
    Total bandwidth per window & 128 MHz\\
    Number of spectral channels per window & 64 \\
    Frequency resolution & 2 MHz \\
    Correlations & RR, RL, LR, LL \\
    \enddata
\end{deluxetable}
\subsection{Data Calibration} \label{sec:calibration}
We used the Common Astronomy Software Applications \citep[\textsc{casa;}][]{2022PASP..134k4501C}, version 6.5, for our data reduction and followed the calibration strategy used by CHANG-ES \citep{2012AJ....144...44I}. For a consistent data calibration across the entirety of the two observations, we opted to manually calibrate the data as opposed to utilizing the \textsc{casa} calibration pipeline. This is primarily for the extensive flagging required due to RFI on the $C$-band data range. Additionally, polarization calibration is not currently part of the automated pipeline routines. We calibrated for the total intensity from the Stokes I continuum, which is the sum of the time-averaged correlations across the right and left circular feeds ($I=\frac{RR+LL}{2}$). Stokes Q and U continuua measure the linearly polarized emission from the cross-hand correlations, RL and LR ($Q=\frac{RL+LR}{2}$, $U=\frac{RL-LR}{2i}$). Stokes V measures the circularly polarized continuum from the difference between the right and left circular correlations  ($V=\frac{RR-LL}{2}$).

\subsubsection{Stokes I Continuum}
\label{sec:stokes_I}

First, we calculated a priori calibrations for the gain curve (the antenna zenith-angle gain dependence), antenna positions, and weather opacity. These were applied to the primary calibrator prior to calculating an initial delay calibration table with \texttt{gaincal}, which removed antenna-induced phase slopes in frequency. We used \texttt{hanningsmooth} on the delay-corrected data, then applied initial flags with \texttt{flagdata} using the automated flagging routine \texttt{tfcrop} on the calibrator sources. Next we used \texttt{setjy} to set the flux density of the primary calibrator. To maintain a consistent model for the polarization calibration, we manually inserted a model for {3C~286} rather than use the built-in model, as the Perley-Butler 2017 model does not model for Stokes Q, U or V.
We scaled the spectral flux density as a function of frequency using the polynomial and coefficients for {3C~286} in \cite{2017ApJS..230....7P} by fitting data between $3.5-14$~GHz. We next calibrated for the amplitude and phase as a function of frequency on {3C~286} using \texttt{bandpass}. With the bandpass applied, we determined the phase and amplitude gain calibrations for all three calibrator sources using \texttt{gaincal}. We bootstrapped the flux amplitude scaling from {3C~286} to the other two calibrators using \texttt{fluxscale}. We then applied the bandpass, gain phase, and amplitude calibrations to the calibrator sources for imaging. Finally, the bandpass and complex gain solutions were applied to the target source from the phase calibrator. After the calibration tables were applied, we carefully checked the visibility data for all calibrators and the target using \texttt{plotms} to remove frequency- and time-based RFI spikes in amplitude and phase. The calibration procedure was then repeated until there was no obvious RFI in the visibility amplitudes.

\subsubsection{Polarization Calibration}

For the polarization calibration, we started by running the automated flagging routines in \texttt{flagdata} (\texttt{tfcrop} and \texttt{rflag}) to remove any residual RFI on the calibration sources. We derived the polarization properties from the NRAO polarization data tables for {3C~286} from 2019 to scale the polarization fraction as a function of frequency.\footnote{\url{https://science.nrao.edu/facilities/vla/docs/manuals/obsguide/modes/flux-density-scale-polarization-leakage-polarization-angle-tables}} The intrinsic polarization angle of {3C~286} for the linearly polarized emission was set to $33\degr$ for the $C$-band observation and $34\degr$ for the $X$-band observation. The coefficients for the flux density, the polarization fraction, and the polarization angles were then input into \texttt{setjy} with their respective reference frequencies for $C$-band (6 GHz) and $X$-band (10 GHz) in order to model the polarization properties of the {3C~286}.

Polarization calibration tables were calculated in the following order. First, the delays between the cross-hand polarizations were calculated with \texttt{gaincal}. Next, we solved for the polarization leakage terms using \texttt{polcal}. The leakage terms are calculated from the zero polarization calibrator and are used to determine instrumental polarization. The polarization angle of 3C 286 was calibrated with \texttt{polcal} to correct for any offset against the model. Finally, the polarization calibration tables were applied to the polarization calibrator targets to verify our proper calibration. We inspected the calibrated phase vs. frequency for both the polarization angle and the polarization leakage calibrators. After calibration, the polarization leakage calibrator showed no phase alignments in the cross-hands which is expected for this unpolarized source. The polarization angle calibrator showed a difference of twice the applied polarization angle in the cross-hands, which is also expected as this is the $R-L$ or $L-R$ phase difference. 
We also imaged the Stokes I, Q, U and V continuua for the polarization angle calibrator in chunks of four spectral windows (512 MHz) to check its polarization angle and fraction. The polarization angle of the electric vector $\theta_{\rm E}$ was then calculated using $\theta_{\rm E}=0.5\arctan\frac{U}{Q}$ (see Section \ref{sec:polmap}). Here, we confirmed that the polarization angles of the electric field vectors on the source were within $33\pm1\degr$ and $34\pm1\degr$ for $C$-band and $X$-band respectively for each image.

Finally, the polarization calibration tables were applied to the target source. The target was separated into its own measurement set, and the task \texttt{statwt} was applied to the target to correct for the weights based on data variance.

\subsection{Imaging} \label{sec:imaging}
\begin{figure*}[h!]
    \centering
    \includegraphics[width=\textwidth]{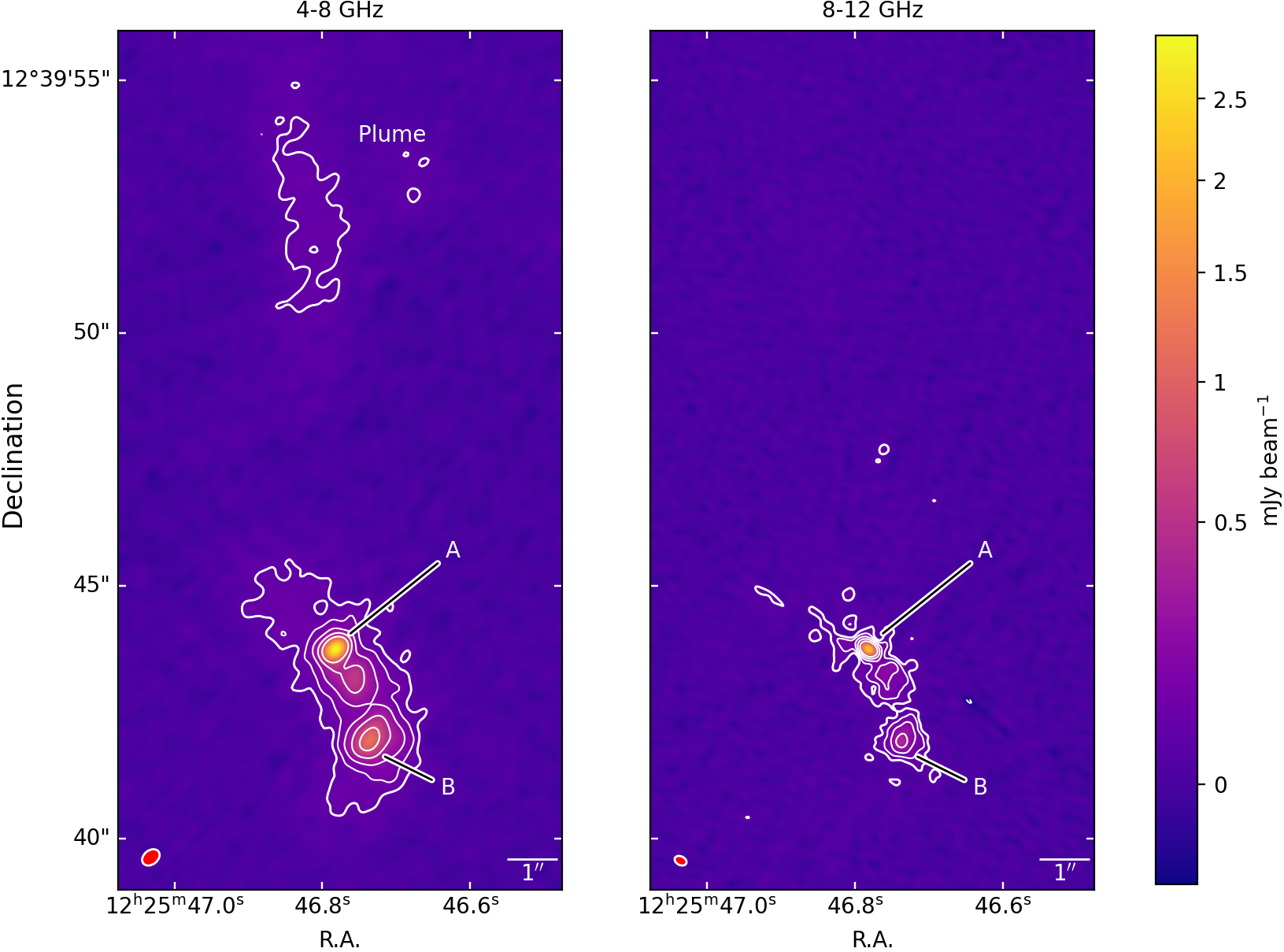}
    \caption{4-8 GHz image of NGC~4388, generated with the MTMFS deconvolution method in \textsc{casa}. Contours are at levels of $(-5,5,10,20,40,80)\times\sigma_I$.
    $1''$ spans $81$ pc at a distance of 16.8 Mpc.
    }
    \label{fig:ngc4388_main}
\end{figure*}
\begin{figure*}[h!]
    \centering
    \includegraphics[width=\textwidth]{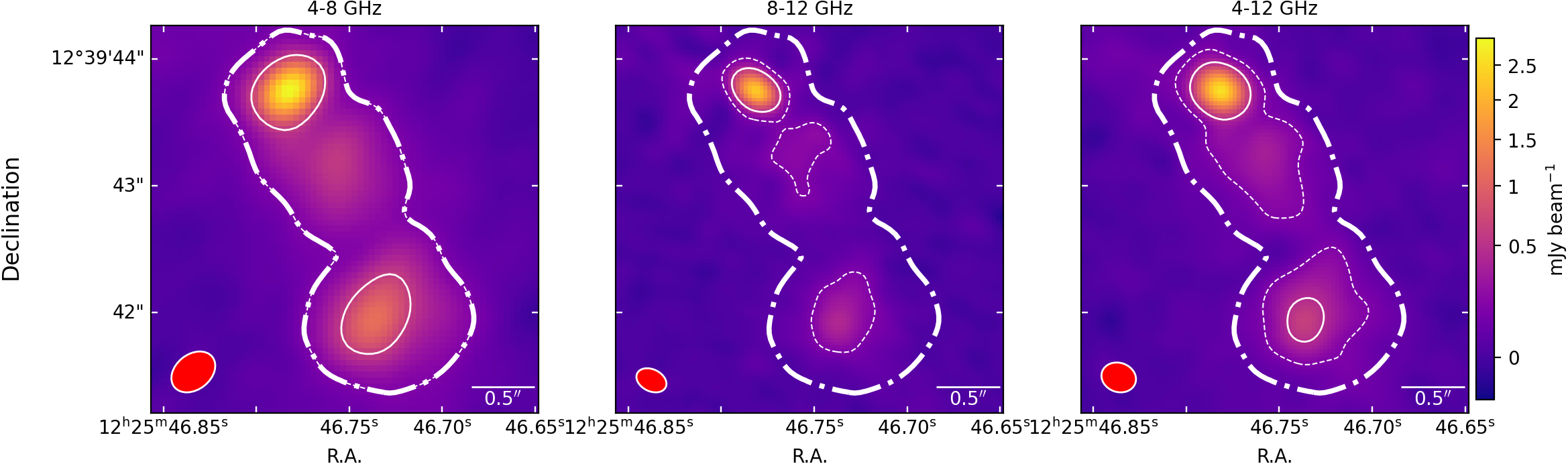}
    \caption{Zoomed in view of the central region of NGC~4388 for the MTMFS images. Left: $C$-band. Middle: $X$-band. Right: $C+X$-band with 7 Taylor terms. The solid white contour lines represent flux densities measured above $60\sigma_{\rm I}$ of the respective image and the dashed lines represent flux densities measured above $15\sigma_{\rm I}$. The thick dash-dot lines represent flux densities above $15\sigma_{\rm I,C-band}$ from the $C$-band image. Measurements are noted in Table \ref{tab:sensitivity}.}
    \label{fig:ngc4388_contours}
\end{figure*}

We imaged the calibrated data using the multi-term, multi-frequency synthesis (MTMFS) imaging algorithm \citep{2011A&A...532A..71R} in \texttt{tclean} with \textsc{casa}. While the standard \textsc{CLEAN} imaging algorithm synthesizes an image describing the spatial structure of the target at a particular frequency, we wanted to utilize the full breadth of these wideband observations in order to also describe the spectral structures of the target. The spatial aspect of the MTMFS algorithm is a multi-scale deconvolution method \cite[described in detail in][]{2008ISTSP...2..793C} which is parameterized as a linear combination of increasing pixel sizes. The spectral structure is parameterized in MTMFS by modeling each flux component as a power law with curvature at some reference frequency, $\nu_0$:
\begin{equation}
    I_{\nu}=I_{\nu_0}\left(\frac{\nu}{\nu_0}\right)^{I_{\rm\alpha}+I_{\rm\beta}\log\frac{\nu}{\nu_0}}.
    \label{eqn:spectral_structure}
\end{equation}
Here, $I_{\nu}$ is the distribution of sky brightness and $\nu$ is the observed frequency. $I_{\rm{\alpha}}$ and $I_{\rm{\beta}}$ are representative of the average spectral index and spectral curvature, respectively. The algorithm works by performing a Taylor expansion on Equation \ref{eqn:spectral_structure} to generate Taylor terms and their associated intensity images. The final MTMFS image is modeled as a linear combination of the images for each spatial scale, $s$, and images for each Taylor-coefficient, $t$:
\begin{equation}
    I_{\nu}^{\rm m}=\sum_{t=0}^{N_t}\sum_{s=0}^{N_s}\left(\frac{\nu-\nu_0}{\nu_0}\right)^t\left[I_s^{\rm shp,\delta}\star I_{s,t}^{\rm sky}\right].
\end{equation}
Here, the outer summation contains $N_t$ Taylor terms and the inner summation over $N_s$ combines each spatial scale. $I_s^{\rm shp,\delta}$ is a tapered, truncated parabola of width proportional to the associated discrete scale, while $I_{s,t}^{\rm sky}$ is a collection of $\delta$-functions describing the intensity of the image. The $\star$ denotes a convolution.

A multi-pronged approach was taken to produce the images of NGC~4388 shown throughout this paper. First, the calibrated $C$-band and $X$-band measurement sets were individually imaged for the Stokes I images, with $1024$ pixels per side and a pixel size set to Nyquist-sample the synthesized beam ($0\farcs046$ for $C$-band and $0\farcs031$ for $X$-band). We used a Briggs weighting scheme with a robustness factor of $0.5$ to provide a good trade-off between resolution and sensitivity for these wideband observations, and cleaned down to a 3$\sigma$ rms stopping threshold. Figure \ref{fig:ngc4388_main} shows the results of imaging with these parameters using MTMFS for our observations, with the $C$- and $X$-band images on the left and right, respectively. In Figure \ref{fig:ngc4388_contours} we show a zoomed-in view of NGC~4388 for the $C$- and $X$-band images for detail. Additionally, a combined $C+X$-band image is also shown in the map on the right, utilizing our widest possible bandwidth (the pixel size in the combined image is the same as the $X$-band image). For the $C$- and $X$-band images, we used the scales $[0,5,10,20,40]\times\rm{\texttt{pixelsize}}$, while for the $C+X$-band image we used an additional scale of size 100.\footnote{For all MTMFS images, we used a small-scale bias of $0.6$, where the small-scale bias parameter weights the largest peak in each scale by multiplying it by $1-\frac{\rm{\texttt{smallscalebias}}}{\rm{\texttt{maxscale}}}$}

We computed 4 Taylor terms for all of our MTMFS images, but also generated an image with 7 taylor terms for the $C+X$-band results. By computing an image with more than one Taylor term, \textsc{casa} generates a spectral index map, i.e.:
\begin{equation}\label{eqn:alpha}
    I_{\rm\alpha}=\frac{I_1}{I_0},
\end{equation}
where $I_0$ and $I_1$ are the intensity maps for the first and second Taylor terms. Additionally, a spectral index uncertainty map $\sigma_{I_{\alpha}}$ is generated by propagating the uncertainty from the ratio of the two Taylor terms used in Equation \ref{eqn:alpha}. In the calculation of this map, smoothed residual maps for each corresponding Taylor term are used as the absolute error. As noted in the \textsc{casa} documentation on Synthesis Imaging\footnote{\url{https://casadocs.readthedocs.io/en/stable/notebooks/synthesis_imaging.html}}, the spectral index error maps should generally only be used as a guide for determining the spectral index from one region in the map relative to another, and not as an absolute uncertainty. When generating an image with three or more Taylor terms we are additionally able to generate a spectral curvature map, where the curvature map is calculated as:
\begin{equation}
    I_{\rm\beta}=\frac{I_2}{I_0}-\frac{I_{\rm\alpha}(I_{\rm\alpha}-1)}{2},
\end{equation}
with $I_2$ being the intensity image from the second Taylor term. No spectral curvature error map is generated when using MTMFS in \textsc{casa}, so we use the spectral index error map as a relative guide. Generally, the spectral curvature errors are proportionally higher than that of the spectral index errors, but we can reduce these errors by introducing many Taylor terms with MTMFS, as we do in this work \citep{2011A&A...532A..71R}. Figure \ref{fig:spix} depicts in-band spectral maps calculated as a result of the Stokes I images for the $C+X$-band imaging.

\begin{deluxetable*}{CCCCCCCCC}
    \tablecaption{Sensitivity}
    \centering
    \tablehead{\colhead{Freq. Range} & \colhead{Taylor} & \colhead{Central Freq. }& \colhead{$\sigma_{\rm I}$} & \colhead{$\sigma_{\rm Q}$}  & \colhead{$\sigma_{\rm U}$}  & \colhead{$\sigma_{\rm V}$}  & \colhead{Restoring Beam} & \colhead{Beam Angle} \\
               ($\rm GHz$)    & \colhead{Terms}       & ($\rm GHz$)             & \multicolumn{4}{C}{($\rm{\mu Jy~beam^{-1}}$)}                                                                                                      & ($\alpha\times\delta$)    & ($\rm{deg}$)}
    \colnumbers 
    \startdata
	\multicolumn{9}{c}{By Frequency Band}\\
	\hline
	4-8  & 4 & 6.1267 & 9.98 & \nodata & \nodata & \nodata & 0.382\arcsec \times 0.278\arcsec & -51.599\degr\\
	8-12 & 4 & 9.9985 & 7.84 & \nodata & \nodata & \nodata & 0.250\arcsec \times 0.172\arcsec & 63.742\degr\\
	4-12 & 4 & 7.9986 & 5.43 & \nodata & \nodata & \nodata & 0.277\arcsec \times 0.229\arcsec & 70.656\degr\\
	4-12 & 7 & 7.9986 & 7.51 & \nodata & \nodata & \nodata & 0.277\arcsec \times 0.229\arcsec & 70.656\degr\\
	\hline
	\multicolumn{9}{c}{By Baseband}\\
	\hline
	4-6    & 4 & 5.1268 & 14.48 & 10.36 & 10.56 & 10.18 & 0.453\arcsec \times 0.350\arcsec & -48.975\degr\\
	6-8    & 4 & 6.9987 & 10.28 & 8.91 & 8.84 & 8.75 & 0.342\arcsec \times 0.244\arcsec & -52.519\degr\\
	8-10   & 4 & 8.9986 & 9.89 & 9.52 & 9.75 & 9.34 & 0.275\arcsec \times 0.189\arcsec & 66.354\degr\\
	10-12  & 4 & 10.9985 & 11.79 & 11.69 & 11.72 & 11.50 & 0.224\arcsec \times 0.155\arcsec & 64.778\degr\\
	\hline
    \enddata
    \tablecomments{Sensitivity measurements for MTMFS images used throughout this paper.\\
    \textbf{Column 1.} frequency band used for imaging. \textbf{Column 2}. number of Taylor terms used. \textbf{Column 3.} center frequency of the image. \textbf{Columns 4--7.} rms \textbf{Columns 8 \& 9.} restoring beam major and minor axis and its position angle.}
    \label{tab:sensitivity}
\end{deluxetable*}

To complement the spectral results from the MTMFS imaging, we test the fidelity of their results by comparing to the spectral energy distributions (SEDs), shown in Figure \ref{fig:sed} for the two bright radio knots in NGC~4388. Here, each 128 MHz spectral window was imaged with the standard CLEAN algorithm with a constant beam at the native resolution of the $C+X$-band MTMFS image. We also applied a uniform \textit{uv}-taper across all spectral windows to create a uniform point-spread function across the full 4 GHz band. In doing so, we apply a uniform spatial scale with a radius of $0\farcs25$ ($20~{\rm pc}$) to make a proper comparison of the flux densities for all spectral windows and in comparison to the MTMFS imaging results for verification of the spectral map fitted results.

\begin{deluxetable*}{CcCCCCCCCC}
    \centering
    \tablecaption{Stokes I Flux Densities and Spectral Properties}
    \tablehead{
    \colhead{Freq. Range} & 
    \colhead{Taylor} &
    \colhead{Center Freq.} & 
    \colhead{$S_{\rm int}$} & 
    \colhead{$S_{\rm peak}$} &
    \colhead{$\log L_{\rm int}$} & 
    \colhead{$S_{\rm peak}/S_{\rm int}$} & 
    \colhead{$\bar\alpha$}   & 
    \colhead{$\bar{\sigma_{\alpha}}$} & 
    \colhead{$\bar\beta$} \\
    (${\rm GHz}$) &           
    \colhead{Terms} &
    (${\rm GHz}$) &           
    ($\rm mJy$)   &           
    ($\rm mJy~beam^{-1}$) &   
    ($\rm erg~s^{-1}$) &   
    &                       
    &                       
    &                       
    }
    \colnumbers                     
    \startdata
	\multicolumn{9}{c}{Component A (within $60\sigma_{\rm I}$ contour)}\\
	\hline
	4-8    & 4 & 6.1267  & 3.72 \pm 0.26^{\dagger} & 2.88 \pm 0.17^{\dagger} & 36.89 & 0.77 \pm 0.07 & -0.02 & 0.02 & -0.09 \\
	8-12   & 4 & 9.9986  & 2.85 \pm 0.15^{\dagger} & 2.29 \pm 0.12^{\dagger} & 36.98 & 0.80 \pm 0.06 & -0.24 & 0.02 & -0.21 \\
	4-12   & 4 & 7.9986  & 3.25 \pm 0.18^{\dagger} & 2.55 \pm 0.13^{\dagger} & 36.94 & 0.79 \pm 0.06 & -0.27 & 0.02 & -0.26 \\
	4-12   & 7 & 7.9986  & 3.22 \pm 0.18^{\dagger} & 2.56 \pm 0.13^{\dagger} & 36.94 & 0.79 \pm 0.06 & -0.28 & 0.04 & -0.32 \\
	\hline
	\multicolumn{9}{c}{Component B (within $60\sigma_{\rm I}$ contour)}\\
	\hline
	4-8    & 4 & 6.1267  & 3.66 \pm 0.22^{\dagger} & 1.14 \pm 0.07^{\dagger} & 36.88 & 0.31 \pm 0.03 & -0.59 & 0.08 & 0.51 \\
 	8-12   & \nodata & \nodata & \nodata & \nodata & \nodata & \nodata & \nodata & \nodata & \nodata \\
	4-12   & 4 & 7.9986  & 2.37 \pm 0.15^{\dagger} & 0.61 \pm 0.04^{\dagger} & 36.81 & 0.26 \pm 0.02 & -0.90 & 0.05 & 0.10 \\
	4-12   & 7 & 7.9986  & 2.04 \pm 0.11^{\dagger} & 0.63 \pm 0.03^{\dagger} & 36.74 & 0.31 \pm 0.02 & -1.07 & 0.11 & 0.60 \\
	\hline
	\multicolumn{9}{c}{Central Total (within $15\sigma_{\rm I,C-band}$ contour)}\\
	\hline
	4-8     & 4 & 6.1267  & 10.62 \pm 0.53 & \nodata & 37.34 & \nodata & \nodata & \nodata & \nodata \\
	8-12    & 4 & 9.9986  & 7.41 \pm 0.38 & \nodata & 37.30 & \nodata & \nodata & \nodata & \nodata \\
	4-12    & 4 & 7.9986  & 8.83 \pm 0.44 & \nodata & 37.37 & \nodata & \nodata & \nodata & \nodata \\
	4-12    & 7 & 7.9986  & 8.84 \pm 0.44 & \nodata & 37.47 & \nodata & \nodata & \nodata & \nodata \\
	\hline
    \enddata
    \tablecomments{Flux density and spectral measurements for MTMFS images used throughout this paper.\\
    $\dagger$ indicates value measured by 2D Gaussian fitting within region selection. All uncertainties contain a 5\% systematic error added. \textbf{Column 1}. frequency band used for imaging. \textbf{Column 2}. number of Taylor terms used. \textbf{Column 3}. center frequency of the image. \textbf{Columns 4--5.} Stokes I integrated and peak flux densities. \textbf{Column 6.} peak luminosity. \textbf{Column 7.} Peak-to-integrated flux density ratio \textbf{Column 8--10.} Mean spectral index, spectral index error and spectral curvature within the FWHM of the fitted region for $S\propto\nu^{\alpha+\beta\log\nu}$.}
    \label{tab:measurements}
\end{deluxetable*}

\subsection{Polarization Maps}
\label{sec:polmap}
To generate the polarization maps we imaged all Stokes parameters simultaneously. The linear polarization intensity map is the result of the combined Stokes Q and U maps, calculated as $P_{\rm{lin}}=\sqrt{I_Q^2+I_U^2}$. Similarly, the total polarization maps are calculated as $P_{\rm{tot}}=\sqrt{I_Q^2+I_U^2+I_V^2}$. Next, the polarization angle maps of the electric vectors were calculated as:
\begin{equation}
\theta_{E}=\frac{1}{2}\arctan\frac{I_U}{I_Q},
\end{equation}
The polarization angle maps of the magnetic field vectors, $\theta_{B}$, is calculated by rotating $\theta_E$ by $90\degr$. Finally, a map of the polarization as a fraction of the Stokes I intensity, $F=\frac{P}{I_I}$, is also generated.

Figure \ref{fig:pol_bb} shows polarization calibration and imaging results for NGC~4388, with the Stokes I intensity map in the top row showing vector lines depicting the polarization angle of the magnetic field. The polarization fraction maps are in the bottom row. The columns show polarization results for each 2 GHz baseband used during the observations (i.e. $4-6$ GHz, $6-8$ GHz, $8-10$ GHz, and $10-12$ GHz) to show polarization features across the $4-12$ GHz spectrum.

\subsection{Error Measurements and Resolved Morphology Considerations}

Table \ref{tab:sensitivity} shows the uncertainty for each stokes intensity map, measured by calculating the rms from a large region in the field off-source. For point source measurements, we used \texttt{imfit} to fit a 2D Gaussian and used the calculated error combined with a 5\% systematic uncertainty for the error on the flux density of the point source. Table \ref{tab:measurements} shows the measured flux densities and the spectral features. The spectral features were measured by the mean value of the pixels within the full-width half-max (FWHM) of the 2D Gaussian fit.

The uncertainty of the linear polarization of the fitted regions were calculated as:
\begin{equation}
\sigma_{P_{\rm lin}}=\sqrt{\left(\frac{I_Q}{P_{\rm{lin}}}\right)^2\sigma_{Q}^2+\left(\frac{I_U}{P_{\rm{lin}}}\right)^2\sigma_{U}^2},
\end{equation}
and are combined with a 5\% systematic uncertainty. The total uncertainty for the polarization angle of their electric vectors are then calculated as:
\begin{equation}
\sigma_{\theta}=\frac{1}{2}\left[\frac{\sigma_{QU}}{1+\left(\frac{I_U}{I_Q}\right)^2}\right],
\end{equation}
where $\sigma_{QU}=\left|\frac{I_U}{I_Q}\right|\sqrt{\left(\frac{\sigma_Q}{I_Q}\right)^2+\left(\frac{\sigma_U}{I_U}\right)^2}$ is the propagated error of $I_U/I_Q$. The fractional uncertainty is similarly calculated as the propagated uncertainty of $P_{\rm lin}/I_I$. Table \ref{tab:pol_measurements} shows the polarization measurements and their uncertainties.

Because there are resolved morphological features in NGC~4388 (and other FRAMEx targets), we also account for systematic contributions of error when measuring the integrated flux densities. We follow \cite{2001ApJS..133...77H} when measuring these uncertainties, where $\sigma_S=\sqrt{N\times\sigma^2+\left(0.05\times S\right)^2}$ is the uncertainty on the flux density of the source, $S$. Here, $N$ is the number of beam areas within the measurement region of the source and $\sigma$ is the rms value from the map.

\section{Results} \label{sec:results}

\subsection{Morphology} \label{sec:morphology}
NGC~4388 is a well-studied object and is known to have a two-component radio
structure in its nuclear region \citep{1987PhDT.........2W,1988ApJ...330..105S}. Figure \ref{fig:ngc4388_main} denotes the northern, brightest radio knot as Component A and the dimmer radio knot to the southwest as
Component B, both of which are mostly unresolved. The detailed view in Figure \ref{fig:ngc4388_contours} depicts the central region containing both components. We consider the ``central region'' to be the structure within the 15$\sigma$ dashed-blue contour lines depicted in Figure \ref{fig:ngc4388_contours}. A complex structure
coupling Components A and B can be seen which weakly detaches from Component
B with increasing frequency.

For the measurements in Table \ref{tab:measurements}, we used a threshold of $60\sigma_{\rm rms}$ to form a region on Components A and B and subsequently fit a 2D Gaussian on the region to determine its peak and integrated flux densities using \texttt{imfit}. There was no $60\sigma$ detection for Component B in the $X$-band image so we did not include any data for it in the table, but there is still flux detected within the $15\sigma_{\rm I,X-band}$ dashed-blue contours. From peak to peak on the $C+X$-band image, Component A has a projected separation from Component B of $1\farcs8$ ($0.15$ kpc). We estimated the combined flux contribution from Components A and B to the central region (dashed-blue contours) and find that radio knots account for $69\pm6\%$ of the central total. Directly to the north of Component A is a plume of diffuse radio emission clearly visible in the $C$-band that is largely resolved away or fainter in the $X$-band image. We measured the integrated flux density of the plume above a $5\sigma_{\rm I,C-band}$ level and find it to be $1.4\pm0.1~{\rm mJy}$ for the $C$-band imaging results.

\subsection{Spectral Structure} \label{sec:spectra}
\begin{figure*}[h!]
    \centering
    \includegraphics[width=\textwidth]{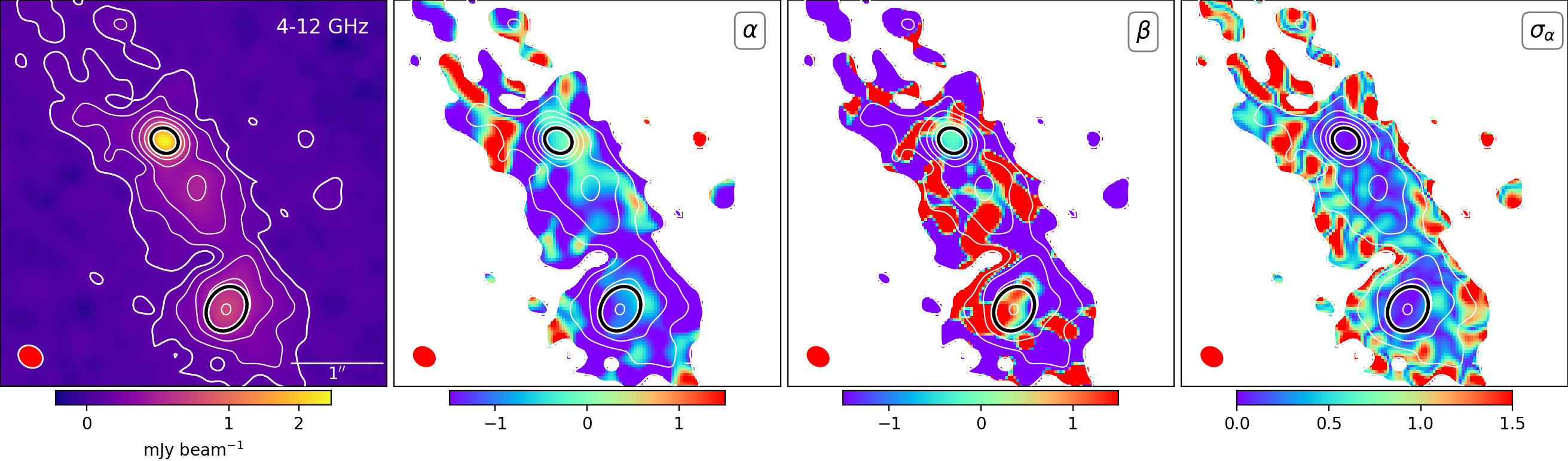}
    \caption{Spectral results of the central region of NGC~4388 resulting from the MTMFS deconvolution of the $C+X$-band imaging with 7 Taylor terms. The white contour lines represent flux densities above $(-5,5,10,20,40,80)\times\sigma_I$. The black demarcations indicate the FWHM of the 2D Gaussian fits. The mean values within the FHWMs are indicated in Table \ref{tab:measurements}. Left: Stokes I intensity map. Middle Left: Spectral index map. Middle Right: Spectral curvature map. Right: Spectral index error map. See Figures \ref{fig:spix_C}--\ref{fig:spix_CX} in Appendix \ref{apx:spix} for the independent $C$- and $X$-band results and the $C+X$-band 4 Taylor term results.}
    \label{fig:spix}
\end{figure*}
\begin{figure*}[h!]
    \begin{center}
    \subfigure[]{\label{fig:sed_compA}\includegraphics[width=\columnwidth]{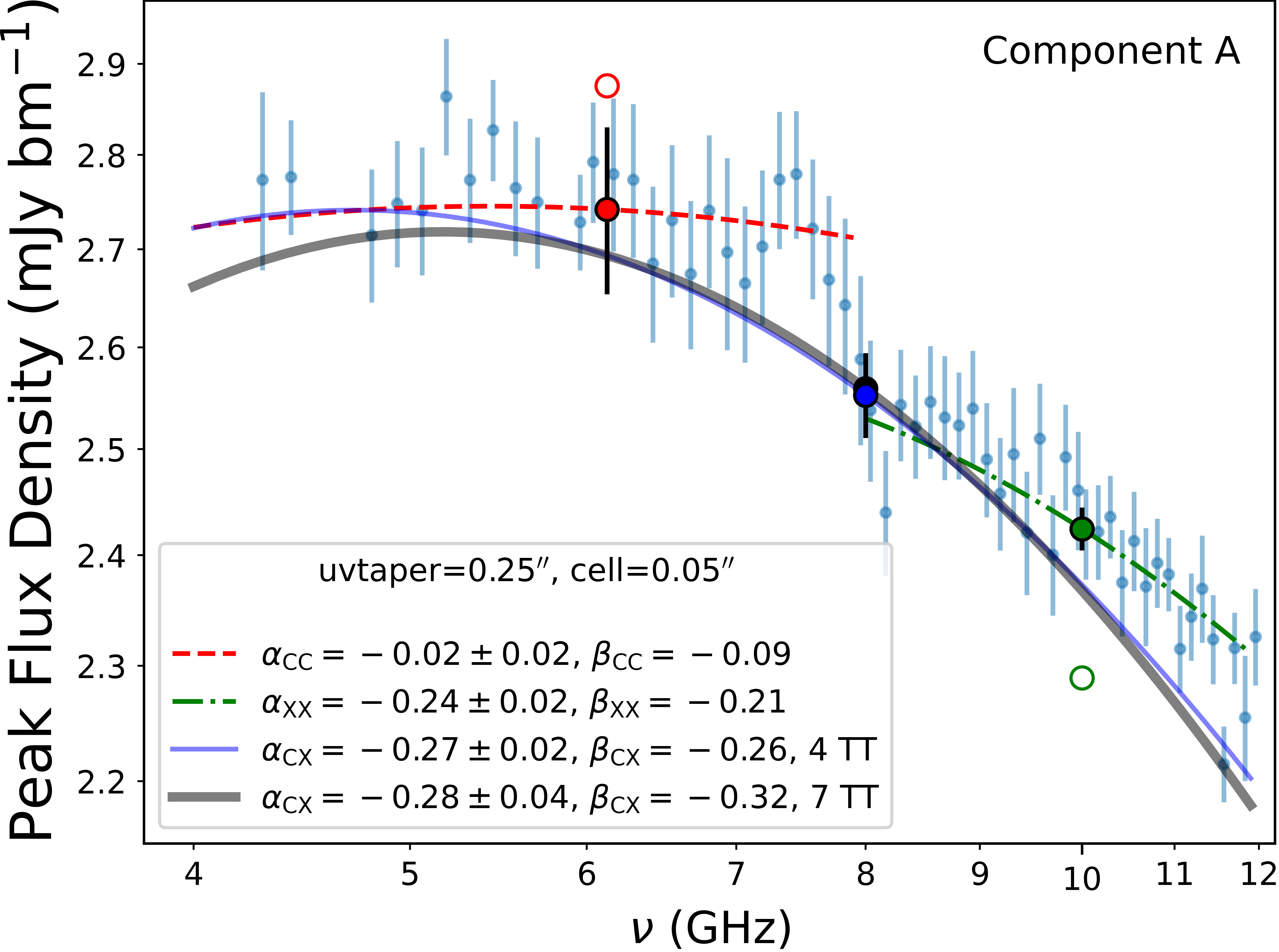}}
    \subfigure[]{\label{fig:sed_compB} \includegraphics[width=\columnwidth]{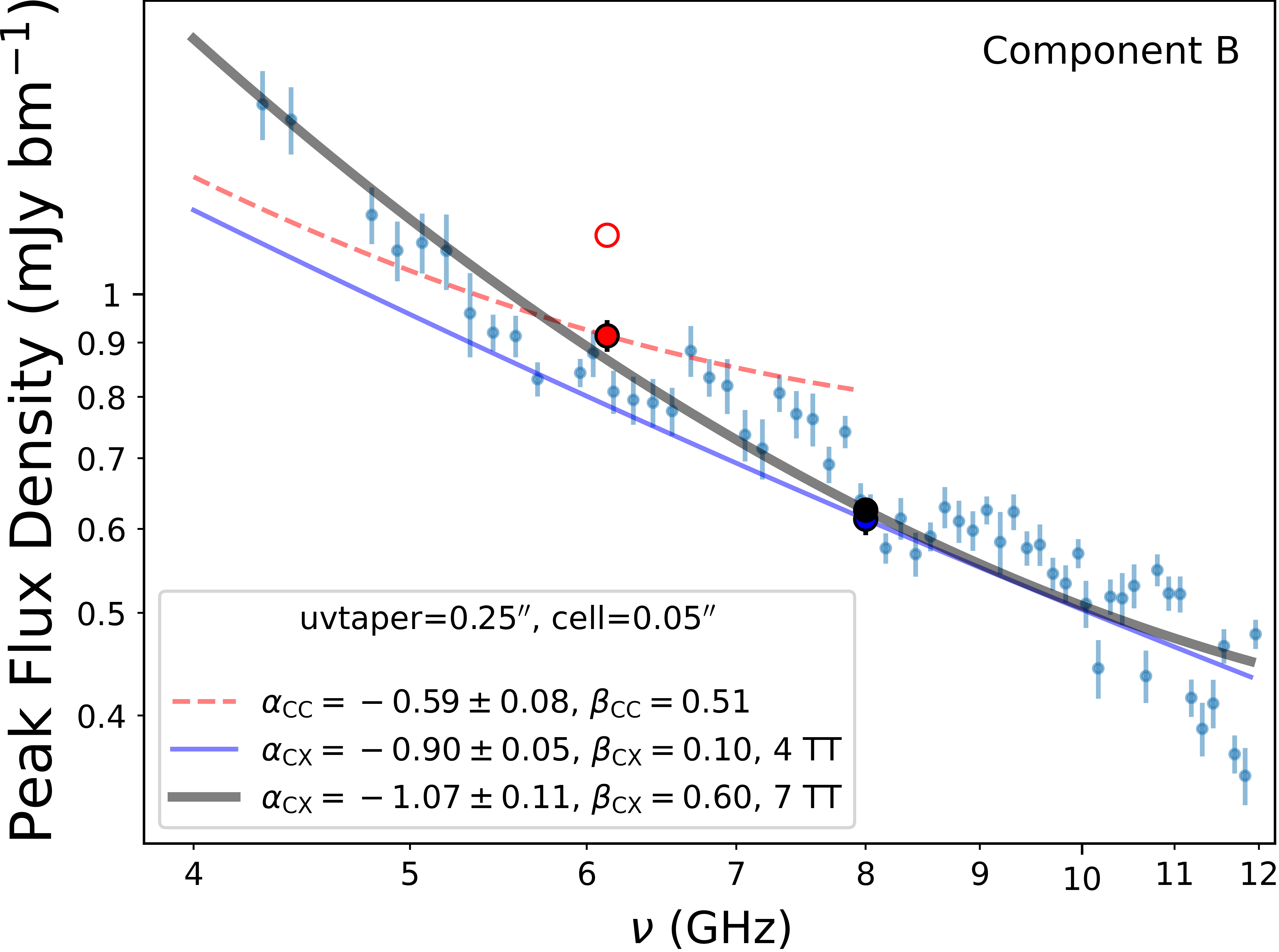}}
	\end{center}
    \caption{\subref{fig:sed_compA} Peak flux densities for Component A across 4-12 GHz. \subref{fig:sed_compB} Peak flux densities for Component B across 4-12 GHz. In both plots, the small blue points are measured from images which span a single 128 MHz spectral window. Each of those images were made with a constant beam set to match the native resolution of the 4-12 GHz MTMFS image and \textit{uv}-tapered with a PSF convolved to $0\farcs25$. The dark blue point represents the peak flux density measurements from the 4 Taylor term $C+X$-band MTMFS imaging runs with the corresponding spectral index and curvature measurements across the bands in the solid blue line. Likewise, the black point and thick solid black line represents the 7 Taylor term $C+X$-band MTMFS imaging runs. The hollow red and green circles and the associated dashed lines represent the peak flux density measurements for the $C$- and $X$-band MTMFS imaging runs, respectively, while the solid circles are shifted to the mean of the channelized data for the corresponding band due to resolution mis-match. No additional 5\% systematic uncertainty error bars are included in these data points for readability.}
    \label{fig:sed}
\end{figure*}
\subsubsection{Spectral Maps}
Figure \ref{fig:spix} shows intensity and spectral maps generated from MTMFS. The spectral maps have distinct components in their morphologies and in Table \ref{tab:measurements} we quote the mean spectral values within the FWHM region of the 2D Gaussian fits of the two radio knots, indicated as the white demarcations.  The Stokes I intensity map is centered at 8.0 GHz, while the spectral index, curvature and error maps depict the spectral information across the combined $4-12$ GHz band. The error map is the uncertainty as a function of position and is sensitive to signal-to-noise. Therefore structures beyond high signal-to-noise regions should not be considered and we have masked out the data that is less than $5\sigma_I$ on the spectral maps.

Component A has a spectral index of $\alpha=-0.28\pm0.04$ within the FWHM region of size $0\farcs31\times0\farcs26$ (1.26 restoring beams) and position angle of $59.7\degr$. Component B has a spectral index of $\alpha=-1.07\pm0.11$ within the FWHM region of size $0\farcs50\times0\farcs41$ (3.27 restoring beams) and a position angle of $166.4\degr$. The mean value of the curvature within the FWHMs for Components A and B are $\beta=-0.32$ and $\beta=0.60$ respectively. Outside of the radio knots, it is unclear how much of the spectral structure is real, as much of the structure is smaller than or close to the size of the restoring beam. It remains possible that there are unaccounted systematic effects causing the structure, especially in the gradients seen across both Components A and B. One possibility is that the structure is due to artifacts from fitting higher order Taylor terms across a broad bandwidth, resulting in an overfitting of the measured data (see Section \ref{sec:sed}).

The spectral maps for $C$-band, $X$-band and $C+X$-band with 4 Taylor terms are shown in the figures in Appendix \ref{apx:spix}.

\subsubsection{Spectral Energy Distribution of Components A and B}
\label{sec:sed}
Figure \ref{fig:sed} shows the channelized data compared to the measured MTMFS spectral results for both Components A and B. Each blue data point indicates a fitted peak flux density for each spectral window. Spectral windows 0, 1, 4, 5, and 14 in the $C$-band observation for NGC~4388 were removed due to RFI. Here, the restoring beam of the channelized data is set to the same size as that of the $C+X$-band images. In all cases, we find that MTMFS is sensitive to spatial resolution. The relative size of the MTMFS restoring beam in comparison to the beam of the channelized data will ``oversample" the flux density if it is larger (as in $C$-band) or ``undersample" it if it is smaller ($X$-band). As we imaged with differing spatial resolutions for the $C$- and $X$-band results, we shifted the flux density amplitudes from their measured values in Table \ref{tab:measurements} to the mean value of the channelized data in their respective bands to see how well the spectral results compare as shown in Figure \ref{fig:sed}.

Component A shows a flat spectrum across $C$-band ($\alpha=-0.02\pm0.02$) and a steeper index with some curvature at $X$-band ($\alpha=-0.24\pm0.02$, $\beta=-0.21$), both of which trace the channelized data well in their respective bands. The two $C+X$-band MTMFS spectral results trace the overall SED shape reasonably well, but diverge at the edges of the bands, especially with the 7 Taylor term image. The poor fit across the entire frequency range may indicate that the data in Component A might be best represented by a broken power law rather than a continuous fit. We applied a least-squares fit to the channelized data to fit a broken power law model using the Levenberg-Marquardt optimization algorithm with \texttt{scipy.optimize.curve\_fit} \citep{2020SciPy-NMeth}. We find a break frequency at $\nu=7.2~{\rm GHz}$ with $\alpha=-0.06$ below the break and $\alpha=-0.34$ above the break. The break frequency aligns with a possible jump in amplitude that peaks around 7.3 GHz.

Component B shows a steep spectrum with curvature in the 7 Taylor term $C+X$-band MTMFS results ($\alpha=-1.07\pm0.11$, $\beta=0.6$) which independently fits the channelized data well across the whole band, but the 4 Taylor term results start to diverge at the lower frequencies. The $C$-band MTMFS results have a steep spectrum ($\alpha=-0.59\pm0.08$, $\beta=0.51$), and again have lower flux densities than the channelized data at the low-frequency edge of the band.

The comparison of MTMFS across our wide 4 and 8 GHz bands to the channelized data shows that MTMFS is not only sensitive to spatial scales but also to bandwidth, at least for these NGC~4388 results. These limitations can potentially be alleviated with higher-order Taylor coefficients at the cost of computation time, and this is suggested as a possibility for increasing the accuracy for the first three Taylor coefficients in \cite{2011A&A...532A..71R}. But the resulting fit from higher-order polynomials is ultimately limited by the theoretical and absolute noise levels of the observation, and may lead to overfitting the data. Lastly, MTMFS generates the spectral maps pixel-by-pixel, but pixel growth methods, such as that used in data cube source finding algorithms \citep[e.g. SoFiA; ][]{2015MNRAS.448.1922S}, may provide more representative sources for fitting spectra across a wide bandwidth.

\begin{figure*}[h!]
    \centering
    \includegraphics[width=\textwidth]{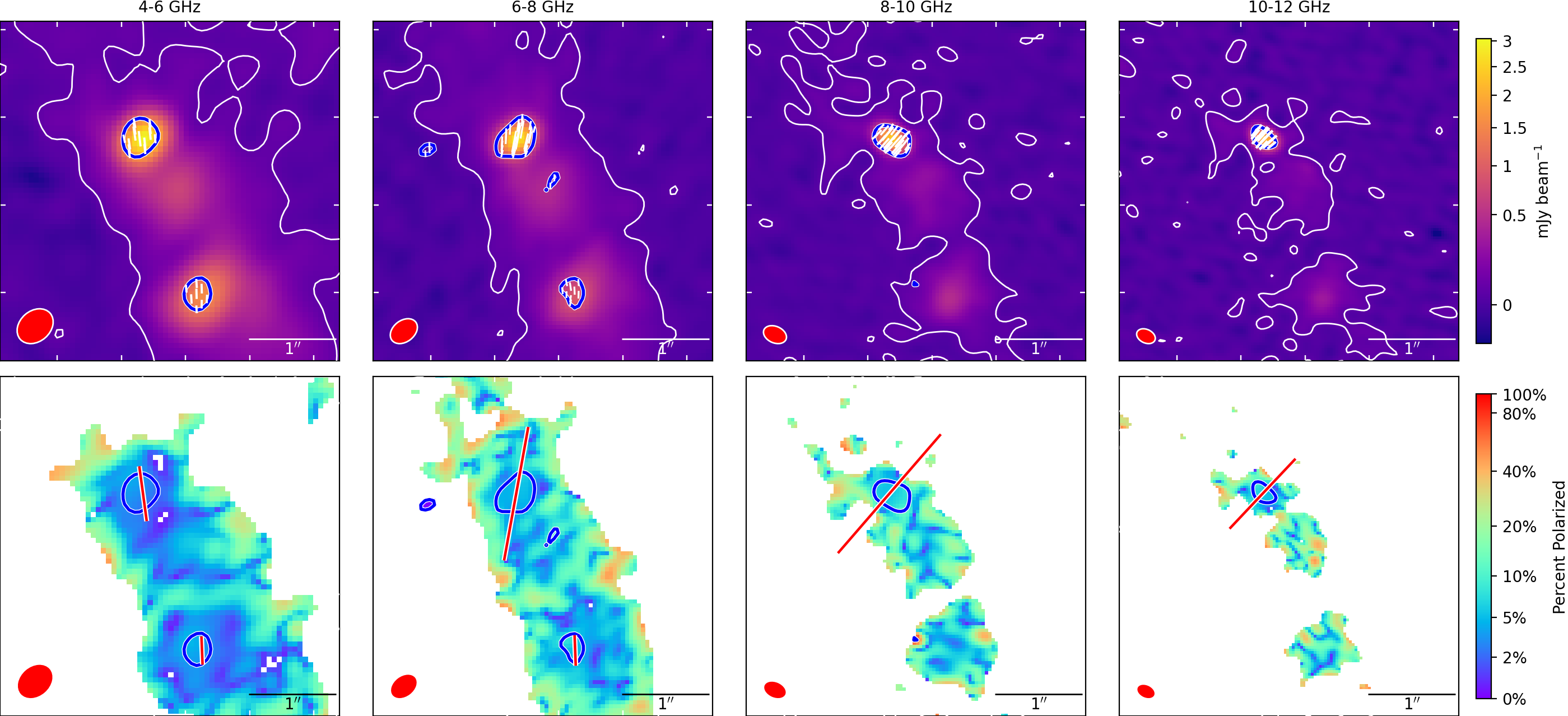}
    \caption{
    Polarization results for NGC~4388 for each 2 GHz baseband. The white contour lines represent a flux density above $5\sigma_I$ from the Stokes I intensity map. The blue contour lines represent linear polarization flux densities above $3\times\sqrt{\sigma_Q^2+\sigma_U^2}$ from the linear polarization map. In the top row, the white vectors indicate the polarization angle of the magnetic field. The bottom row shows the linear polarization as a percentage of Stokes I intensity with the red lines as the polarization angle and relative intensity from Table \ref{tab:pol_measurements}.}
    \label{fig:pol_bb}
\end{figure*}
\begin{deluxetable*}{CCCCCC}[h!]
    \centering
    \tablecaption{Polarization Properties}
    \tablehead{
    \colhead{Freq. Range} & 
    \colhead{Center Freq.} & 
    \colhead{$S_{\rm int,I}$} & 
    \colhead{$P_{\rm lin}$} & 
    \colhead{$P_{\rm lin}/S_{\rm int,I}$}  & 
    \colhead{$\theta_{\rm Pol.}$}\\
    (${\rm GHz}$) &           
    (${\rm GHz}$) &           
    ($\rm mJy$) &             
    ($\rm mJy$) &             
    \% &                    
    ($\rm deg$)             
    }
    \colnumbers                    
    \startdata
	\multicolumn{6}{c}{Component A}\\
	\hline
	4-6      & 5.1268  & 1.66 \pm 0.08 & 0.05 \pm 0.01 & 2.9 \pm 0.5 & 8 \pm 5 \\
	6-8      & 6.9987  & 2.55 \pm 0.13 & 0.12 \pm 0.02 & 4.7 \pm 0.7 & 170 \pm 3 \\
	8-10     & 8.9986  & 2.20 \pm 0.11 & 0.14 \pm 0.02 & 6.3 \pm 1.0 & 139 \pm 3 \\
	10-12    & 10.9985  & 1.48 \pm 0.07 & 0.09 \pm 0.01 & 5.8 \pm 1.0 & 137 \pm 4 \\
	\hline
	\multicolumn{6}{c}{Component B}\\
	\hline
	4-6      & 5.1268 & 0.65 \pm 0.03 & 0.03 \pm 0.01 & 3.8 \pm 0.8 & 2 \pm 8 \\
	6-8      & 6.9987 & 0.49 \pm 0.03 & 0.03 \pm 0.01 & 5.6 \pm 1.2 & 2 \pm 7 \\
	\hline
    \enddata
    \tablecomments{Polarization measurements from a region within the $3\times\sqrt{\sigma_{\rm Q}^2+\sigma_{\rm U}^2}$ contours of the linear polarization map shown in Figure \ref{fig:pol_bb}. All images were generated with 4 Taylor terms.\\
    \textbf{Column 1.} frequency band used for imaging. \textbf{Column 2.} center frequency of the image. \textbf{Column 3.} integrated flux density from the Stokes I intensity map. \textbf{Column 4.} integrated flux density from the linear polarization intensity map. \textbf{Column 5.} per cent polarized. \textbf{Column 6.} polarization angle of region (the direction of the magnetic vectors are ambiguous by $180\degr$).}
    \label{tab:pol_measurements}
\end{deluxetable*}
\subsection{Polarization Results} \label{sec:polarization}
Figure \ref{fig:pol_bb} shows a detailed view of the polarization structure across the central region per baseband. No polarized emission was detected outside of Components A and B and no circular polarization was detected for any of our polarization maps. The sources we detect have polarized emission at the $3-6$\% level across the four basebands. We find that there appears to be Faraday rotation across Component A, indicating a wavelength-dependent depolarization due to magnetized plasma in the foreground, causing a rotation in the polarization angle in the linearly polarized radiation \citep{1966MNRAS.133...67B}. Component B only had a significant detection in the $C$-band images and it is unclear whether Faraday rotation is also affecting Component B as we do not have an $X$-band detection.
Appendix \ref{apx:stokes} shows the resulting Stokes I, Q, U, V and linear polarization maps without any magnetic field vectors overlaid. In a future paper in this series, we intend to apply rotation measure synthesis to all of our VLA FRAMEx targets to correct for depolarization effects \citep[e.g. ][]{2005A&A...441.1217B,2009IAUS..259..591H}.

\section{Discussion} \label{sec:discussion}

\subsection{NGC~4388 in Context}
\label{sec:context}

\begin{figure*}
    \begin{center}
    \includegraphics[width=\textwidth]{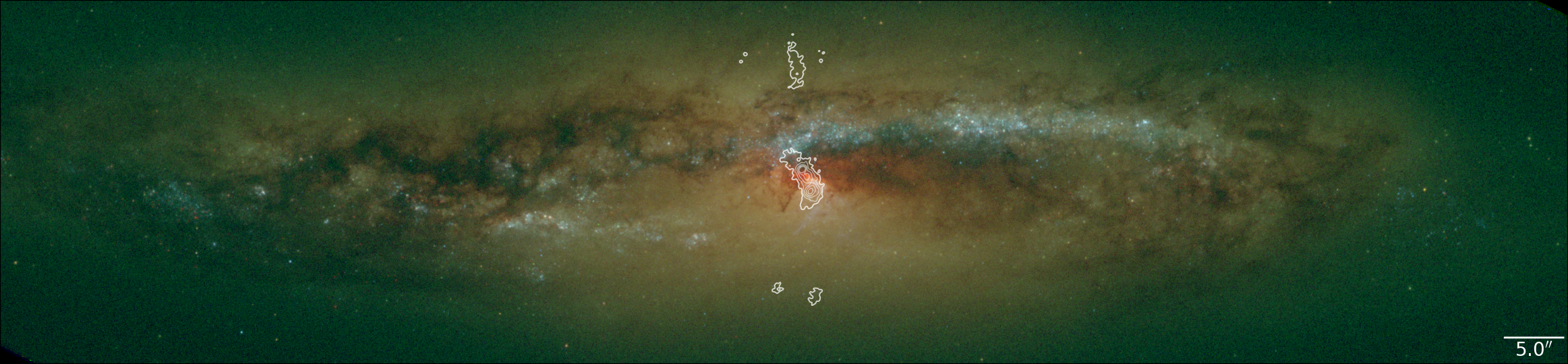}
	\end{center}
    \caption{Hubble WFC3 with F814W (red), F438W (green) and F336W (blue) filters. The white contour lines represent the $C$-band image results shown in Figure \ref{fig:ngc4388_main} at  $(-5,5,10,20,40,80)\times\sigma_I$ levels.}
    \label{fig:ngc4388_context}
\end{figure*}

NGC~4388 is a nearly edge-on spiral galaxy 16.8 Mpc away with a systemic velocity of $2538~{\rm km~s^{-1}}$ \citep{2003MNRAS.345.1057M,2009AJ....138.1741C}. The disk of the galaxy is highly inclined relative to our line of sight (${\sim}77-79\degr$), with the north edge of the galaxy being the nearer side \citep{1988ApJ...332..702P,1990AJ....100..604C,2012AJ....144...44I}. For a useful reference, Figure \ref{fig:ngc4388_context} shows an image from the Hubble Data Archive with our $C$-band contours overlaid (Program ID GO-12185; PI Greene).  \cite{1982MNRAS.199..905P} were the first to show that NGC~4388 was within the Virgo cluster, a collection of galaxies centered on the supergiant elliptical galaxy M87 \citep[$d_{\rm M 87} = 16.4\pm0.5$ Mpc;][]{2010A&A...524A..71B}. While many spiral galaxies within the cluster show \ion{H}{1} deficiency, NGC~4388 has severe \ion{H}{1} truncation relative to its optical disk, particularly on the western side \citep{1990AJ....100..604C,2009AJ....138.1741C}. The truncation is likely a result of ram-pressure stripping as it traverses through hot X-ray gas within the cluster. A large-scale tail of \ion{H}{1} discovered by \cite{2005A&A...437L..19O} extends $110~{\rm kpc}$ to the northeast from NGC~4388, well into the Virgo cluster's central region. They suggest that the ram-pressure stripping is likely due to the nearby M86, another elliptical galaxy with a hot X-ray halo just northeast of NGC~4388. Furthermore, \cite{1988ApJ...332..702P} studied a map of the [\ion{O}{3}]/(H$\alpha$+[\ion{N}{2}]) and found two conical structures that appear to be photoionized: one extends to the northeast and comes into view at the north edge of the disk while the other extends to the southwest within the disk. The northeast cone was later found to extend well outside of the galaxy, forming an extended neutral and ionized structure when combined with the long \ion{H}{1} tail \citep{2002ApJ...567..118Y,2005A&A...437L..19O}.
\begin{figure}
    \begin{center}
    \subfigure[]{\label{fig:xray}\includegraphics[width=\columnwidth]{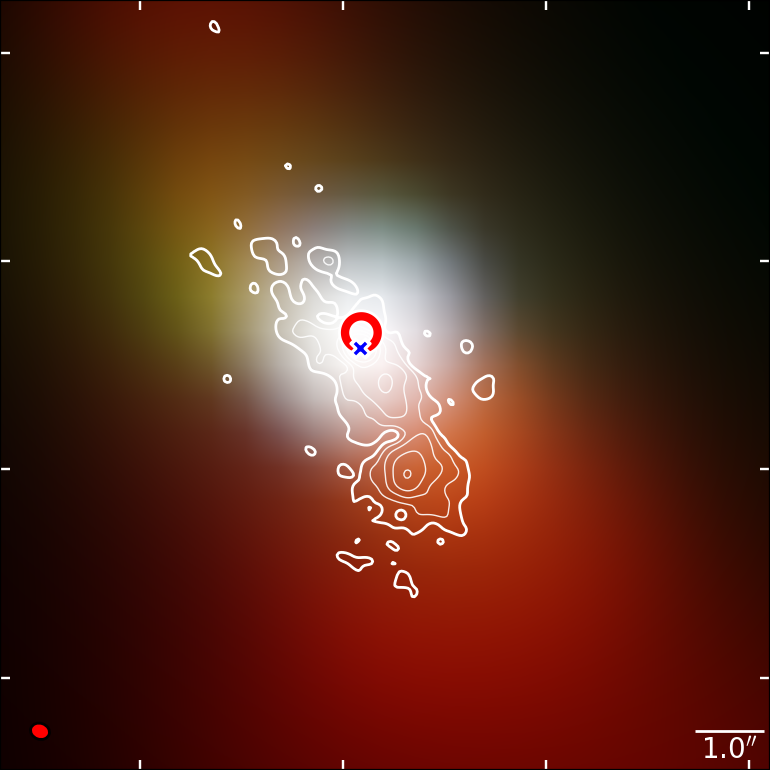}}
    \subfigure[]{\label{fig:OIII}\includegraphics[width=\columnwidth]{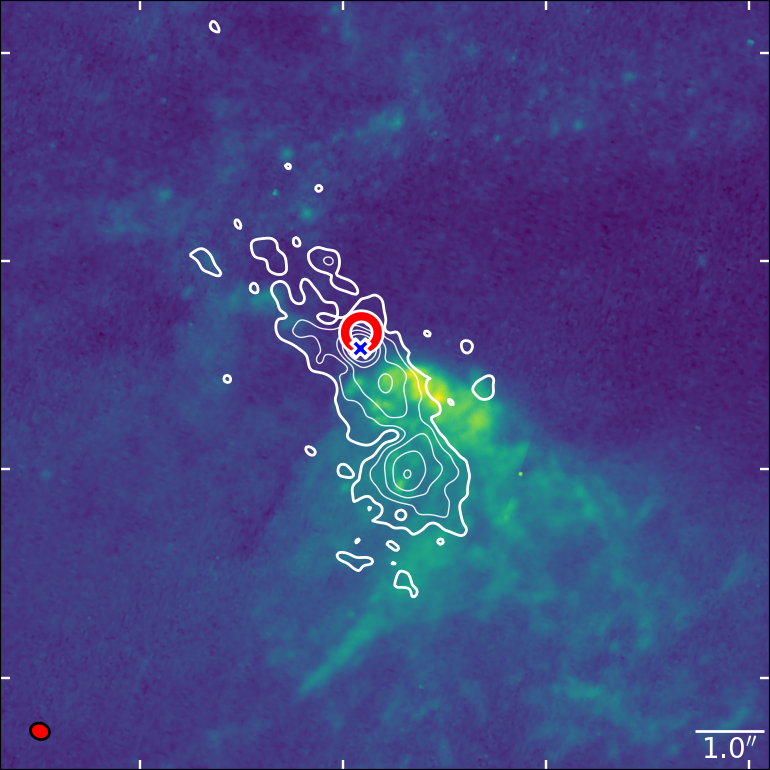}}
    \caption{
    \subref{fig:xray} Adaptively smoothed Chandra X-ray image of NGC~4388 using \textsc{ciao} 4.14 \citep{2006SPIE.6270E..1VF}. The red is $0.2-2$ keV, green is $2-4$ keV and blue is $4-10$ keV. The hard X-ray position is centered at $12^{\rm h}25^{\rm m}46.78^{\rm s}~+12\degr39\arcmin43.97\arcsec$ with a $1\sigma$ uncertainty of $0\farcs24$.
    \subref{fig:OIII} Hubble WFC3 FQ508N image of NGC~4388 showing the [\ion{O}{3}] line emission.
    \textbf{Both:} The white contour lines represent the $C+X$-band image results shown in Figure \ref{fig:ngc4388_contours} for flux densities above $(-5,5,10,20,40,80)\times\sigma_I$. The red, solid circle represents the Chandra X-ray position with an uncertainty of $0\farcs2$. The VLBA detection's position is marked in the blue \textcolor{blue}{\boldmath$\times$} (\citetalias{2022ApJ...936...76S}).}
    \label{fig:multiwavelength}
    \end{center}
\end{figure}

\subsection{The Polarized Radio Continuum}

\cite{2016ApJ...824...30D} observed the disk and halo of NGC~4388 with the VLA in its C- and D-configurations at $5-7$ GHz in an effort to study ram-pressure effects of the polarized radio continuum and magnetic fields. Their observations show two highly polarized outflows directly north and south of the nuclear region where Components A and B reside. Their northern outflow directly aligns with the plume but we detect no polarization in that region. In contrast to the detection of nuclear polarization in our observations, their 6 GHz observations probe down to $2.3~{\rm\mu Jy~beam^{-1}}$ (at a resolution of $5\farcs33\times5\farcs33$) and they do not detect any significant polarization in the nuclear region \citep[see Figure 3 of ][]{2016ApJ...824...30D}. The disparity is likely due to a combination of differences between our observational parameters and array configurations, which result in a difference in resolution compared to that derived in \cite{2016ApJ...824...30D}.

As Component A is viewed through the edge of the disk of the host galaxy, it is potentially subject to depolarization effects along the line of sight \citep[e.g. the background rotation measures modified by \ion{H}{2} bubbles in the ISM of the Galaxy; ][]{2018ApJ...865...65C}. Rotation measure synthesis will enable us to estimate the frequency-dependent depolarization, or Faraday dispersion, and correct for rotation effects across our wideband observations \citep{2005A&A...441.1217B,2009IAUS..259..591H}. We intend to apply rotation measure synthesis to this and other VLA FRAMEx targets in future work, as was done in \cite{2016ApJ...824...30D}, so a direct comparison of the magnetic field vectors may be misleading at specific frequencies.

Without rotation measure synthesis applied, our magnetic field vectors at $C$-band almost directly align with the north-south magnetic field vectors measured across the northern outflow in \cite{2016ApJ...824...30D}, and we can trace the magnetic field vectors from the nucleus all the way to a galactic-wide scale in the halo. The magnetic field vectors measured in \cite{2016ApJ...824...30D} extend a projected $50''$ ($4.1~{\rm kpc}$) in the north along a filament that spans $3\farcs9$ ($0.3~{\rm kpc}$) across with a position angle of $20-30\degr$. Linearly polarized synchrotron emission is generally less affected by Faraday rotation for frequencies ${>}5$ GHz \citep{Beck2013}, so it is possible that this alignment for polarization angles across 4-8 GHz will differ after correcting for it.

\subsection{The AGN in Component A}
\label{sec:component_A}
\cite{1990ApJ...362..434C} were the first to suggest that Component A is likely emission surrounding the AGN due to its flat spectrum. As such, we briefly explore Component A's position relative to detections at various wavelengths. While initial observations of NGC~4388 in \citetalias{2021ApJ...906...88F} showed no detection at VLBI resolution (rms~$\sim108$\,$\mu$Jy), the deeper observation (rms~$\sim20$\,$\mu$Jy) in \citetalias{2022ApJ...936...76S} made a ${\sim}12\sigma$ detection. The detection is centered on Component A with a $5.86$ GHz peak flux density of $0.214\pm0.018~{\rm mJy~beam^{-1}}$, which is $7.4\%$ of our $C$-band peak flux density measurement.

Figure \ref{fig:xray} shows \textit{Chandra X-ray Observatory} (CXO) imaging results (ObsID 1619, 19.97 ks exposure; PI: Wang) for NGC 4388 at 0.2--2 keV, 2--4 keV and 4--10 keV. As hard X-rays reliably identify AGN activity, we determined a statistically significant hard X-ray source from the CXO imaging results. The center of the X-ray source is consistent with Component A with a $1\sigma$ position uncertainty of $0\farcs24$ when a $\sim0\farcs2$ offset from Gaia is accounted for. The X-ray position is shown as the solid red circle in Figure \ref{fig:multiwavelength}. We re-imaged the VLBA deep observation of NGC~4388 from \citetalias{2022ApJ...936...76S} with the phase center set to the coordinates of Component B from this work and found no corresponding detection. We were only able to make a detection coincident with Component A when we imaged with a wider field of view as seen by the VLBA that encompassed Components A and B. Given that there was no detection at VLBI resolution outside of Component A, in combination with the hard X-ray position on top of Component A, it seems likely that Component A contains the AGN.

The SED for Component A in Figure \ref{fig:sed_compA} shows a flat spectrum below the break frequency measured at 7.2~GHz, and a steeper but still relatively flat spectrum above the break frequency. The spectral shape from our VLA results is reminiscent of SEDs for compact jets described in the theoretical work of \cite{1979ApJ...232...34B}. There, an unresolved compact radio source (${<}1~{\rm kpc}$) with a relativistic jet has an SED with a flat spectrum up to some maximum brightness temperature, where the spectral index steepens to $\alpha=-1$ due to synchrotron losses. While we do not have any flux density measurements to continue the SED beyond 12 GHz, \cite{2022MNRAS.515..473P} observed NGC~4388 with the VLA in its B-array configuration at $Ku$-band (15 GHz) and at the C-array configuration at $K$-band (22 GHz) as part of a multi-frequency study of nearby ($0.003 \leq z \leq 0.3$), hard X-ray selected AGNs. They measured a 15 GHz peak flux density of $2.3\pm0.1~{\rm mJy~beam^{-1}}$ with a restoring beam size $0\farcs30\times0\farcs21$ and a 22 GHz peak flux density of $2.6\pm0.1~{\rm mJy~beam^{-1}}$ with a restoring beam size of $0\farcs7\times0\farcs37$ (roughly 1 and 4 times the size of our $C+X$-band beam, respectively). If we extend our power-law spectral fits (Section \ref{sec:sed}) beyond the 7.2 GHz break frequency for Component A, we find that we would expect a 15 GHz peak flux density of $2.0~{\rm mJy~beam^{-1}}$, while for a 22 GHz peak flux density we would expect $1.6~{\rm mJy~beam^{-1}}$, assuming there is no other spectral break or curvature. Though the $Ku$-band flux density is within 0.3 mJy of our spectral fit extension, there is a ${\sim}1~{\rm mJy~beam^{-1}}$ excess when compared to the $K$-band measured flux density. Given the beam size, it is possible that the larger $K$-band restoring beam is capturing more radio emission with respect to our extrapolated expectations, but we cannot rule out radio emission from compact jet processes as described in \cite{1979ApJ...232...34B} based on these results alone.

On the other hand, when zoomed into VLBA sub-pc spatial scales, the results from Shuvo et al. (in prep.; FRAMEx~V) show NGC~4388 as an unresolved point-source with no extended or collimated features across multiple frequencies that would indicate a jet or jet-like activity. The four frequencies observed in FRAMEx~V show that NGC~4388 has a nearly flat band-to-band spectral index of $\alpha=0.12$ across $C$-band ($0.23\pm0.02~{\rm mJy~beam^{-1}}$ at 4.4 GHz) and $X$-band ($0.25\pm0.02~{\rm mJy~beam^{-1}}$ at  8.6 GHz) from the peak flux densities, which are respectively ${<}10\%$ of the the $C$- and $X$-band VLA measurements. Additionally, a detection at $L$-band was found ($0.19\pm0.02~{\rm mJy~beam^{-1}}$ at 1.6 GHz) but there was no detection in the $K$-band observation (22 GHz), providing a $5\sigma$ upper limit of ${<}0.40~{\rm mJy~beam^{-1}}$.

Section 4.2 in FRAMEx~V discusses an anti-correlation between source size and turnover frequency, where younger objects have higher turnover frequencies \citep[also see][]{1997AJ....113..148O}. It remains possible that the ram-pressure stripping activity caused the AGN in NGC~4388 to ``turn on," and estimates put peak ram-pressure stripping around $190-200~{\rm Myr}$ ago \citep{2010A&A...514A..33P,2018A&A...620A.108V}. High-resolution \ion{H}{1} imaging from MeerKAT or future telescopes such as the Square Kilometer Array or ngVLA could provide more precise measurements of the \ion{H}{1} activity and could potentially narrow down the age and source of the emission within the disk of the host galaxy.

\subsection{The Excess VLA Emission}

The detected VLBA flux densities in \citetalias{2021ApJ...906...88F} are all ${<}7$\% than that of archival VLA flux densities, and it was suggested that the extranuclear radio emission from the VLA is likely due to shocks near the nucleus. In the case of NGC 4388, we find that the VLBA flux density at Component A is ${\lesssim}10$\% of that of the VLA measurement, giving a higher VLBA percentage of the VLA flux density than other FRAMEx sources. We explored the excess VLA emission by measuring a brightness temperature for our $C+X$-band results at Component A to be $10^{2.9}~{\rm K}$. This is perhaps indicative of a region of star formation, such as in the \ion{H}{2} regions studied in \cite{1994ApJS...91..659K} which showed brightness temperatures of ${\lesssim}10^4~{\rm K}$ \citep[also see Figure 4 of ][]{1992ARA&A..30..575C}. By contrast, the brightness temperature of $10^{5.8}~{\rm K}$ for Component A from the VLBA measurements in \citetalias{2022ApJ...936...76S} is consistent with a weak AGN \citep[e.g.,][]{2005ApJ...621..123U,2018A&A...619A..48R} or perhaps strong star formation. Thus we calculated the star formation rate (SFR) from the integrated flux density as quoted in Table \ref{tab:measurements} using Equation 15 from \cite{2011ApJ...737...67M}. This equation was calibrated on individual extranuclear \ion{H}{2} star forming regions and accounts for thermal and nonthermal radio emission processes. From this equation we estimate the SFR of Component A to be $0.11\pm0.01~{\rm M_{\odot}~yr^{-1}}$, which is similar when compared to the star forming regions shown in Figure 3 of \cite{2011ApJ...737...67M}, where 9 out of 10 sources have a total SFR of $0.01-0.10~{\rm M_{\odot}~yr^{-1}}$.

As a point of comparison, we also calculated an upper limit on the SFR in the nuclear region of NGC 4388 using the optical tracer H$\alpha$. We used Equation 7 from \cite{2011ApJ...737...67M}, which again was calibrated for star forming regions, to calculate an extinction-corrected SFR from H$\alpha$,
\begin{equation}
\left(\frac{\rm SFR_{mix}}{M_{\odot}~{\rm yr}^{-1}}\right)=5.37\times10^{-42}\left[\frac{L^{\rm obs}_{\rm H{\alpha}}+0.02\cdot\nu L_{\nu}(24{\rm \mu})}{\rm erg~s^{-1}}\right].
\end{equation}
Here, the SFR is determined using a mix of H$\alpha$ luminosity ($L^{\rm obs}_{\rm H\alpha}$), and infrared (IR) luminosity ($\nu L_{\nu}(24~{\rm \mu})$), where the IR luminosity has a correction factor optimized for $\nu L_{\nu}(24~{\rm \mu})<4\times10^{42}~{\rm erg~s^{-1}}$. We used the H$\alpha$ map from CHANG-ES \citep[e.g.][]{2015AJ....150...81W,2019ApJ...881...26V} to fit a 2D Gaussian surrounding the bright, nuclear component and measured an integrated flux density of $F_{\rm H\alpha}=1.05\pm0.18\times10^{-13}~{\rm erg~s^{-1}~cm^{-2}}$. For the IR luminosity, we used \textit{Spitzer} MIPS $24{\rm \mu}$ observations from \cite{2012MNRAS.423..197B} imaging and again fit a 2D Gaussian around the nuclear region to measure an integrated flux density of {$\nu F_{\nu}(24{\rm \mu})=(5.97\pm0.30)\times10^{-12}~{\rm erg~s^{-1}~cm^{-2}}$}. Combining the measurements, we calculated an extinction-corrected SFR from H$\alpha$ of ${\rm SFR}_{\rm mix}=(4.07\pm0.34)\times10^{-2}~{\rm M_{\odot}~yr^{-1}}$, roughly 37\% of the SFR calculated from our Component A radio measurement. Here, the radio emission appears to be too luminous for star formation alone, suggesting that the AGN is contributing to the radio unless there is significant extinction due to dust that is not accounted for.

\subsection{Energy Budget at the Nucleus}
If the SFR calculated in the previous section does not sufficiently contribute to the excess nuclear radio emission, another contributor to the emission is potentially due to AGN winds. 
Figure 5 in FRAMEx~V shows how the accretion state of an AGN, which can be parameterized by a radio-loudness parameter as a function of the Eddington ratio ($L_{\rm bol}/L_{\rm Edd.}$), corresponds to winds or jets being the primary source of radio emission in an AGN. For highly-accreting AGNs, a wind-based scenario is favored, where radio emission is produced by shocked interactions in the dense neighboring ISM if the outflows are strong enough. In this interpretation, due to its low radio-loudness, NGC~4388 falls in the region of highly-accreting, younger AGNs, leaving the wind-based scenario more likely than jet-like emission. In this section, the energy deposited into winds is calculated against the kinetic energy from the synchrotron emission to verify the energy budget for the nucleus.

\subsubsection{Mechanical Feedback}
We followed the method to estimate the mechanical power that was used in \cite{2023ApJ...958...61F} (\citetalias{2023ApJ...958...61F}), where they studied the energy budget for the FRAMEx galaxy NGC~3079. We first calculated the mean total power from the synchrotron emission, $\left<P\right>=\frac{4}{3}\sigma_T\beta^2\gamma^2cU_B$, where $\sigma_T$ is the Thomson cross-section, $\beta=\frac{v}{c}$, $\gamma$ is the Lorentz factor, and $U_B=\frac{B^2}{8\pi}$ is the magnetic energy density. We approximated the Lorentz factor by the equation
\begin{equation}
\gamma\approx\left(\frac{2\pi m_{\rm e}c\nu}{eB}\right)^{\frac{1}{2}},
\end{equation}
where $\nu$ is the observing frequency. We used the estimated magnetic field strength across the nucleus from \cite{2016ApJ...824...30D}, $B_{\rm tot}=67~{\rm \mu G}$, which was corrected for Faraday rotation. We find that the Lorentz factor ranges between 5710 and 7300 for our $C$-band and $X$-band imaging frequencies, which are similar to the Lorentz factors calculated for NGC~3079. By dividing the integrated luminosities for Component A from Table~\ref{tab:measurements} by the mean power, we estimated the required number of electrons ($N_{\rm e}$) to produce the observed radio emission. From there, we calculated the mechanical energy in the nuclear region using ${\rm KE}=2N_{\rm e}m_{\rm e}c^2(\gamma-1)$. This is the relativistic kinetic energy from the synchrotron emission observed, where the factor of 2 comes from assuming that the ISM has an equal number of protons and electrons that may have been shocked and accelerated. We find that for all observing frequencies the kinetic energy is on the order of ${\rm KE}\sim10^{50}~{\rm erg}$, consistent with the low end of the range measured for the bright emission at VLBA scales for NGC~3079 (${\rm KE}\sim10^{50}-10^{52}~{\rm erg}$).

\subsubsection{Radiative Feedback}
\cite{2014MNRAS.442..784Z} proposed that in the absence of a compact jet for their sample of 568 quasars, wind-driven shocks accelerate particles that interact with the ISM to produce the synchrotron emission observed in the radio. They suggested that 4\% of the bolometric luminosity can be converted into the kinetic luminosity of AGN winds. Similarly, \cite{2017A&A...601A.143F} conducted a multi-wavelength study across 94 AGN with multiple redshifts that have detected winds at sub-pc to kpc scales, to develop scaling relationships between AGN activity, radiation, winds, and star formation. They find that the kinetic luminosity from AGN winds is on the order of $1-10\%$ of the bolometric luminosity for molecular winds and $0.1-10\%$ for ionized winds. We used the bolometric correction from the BAT AGN Spectroscopic Survey \citep[BASS; Equation 1 in ][]{2023MNRAS.518.2938T} and calculated a bolometric luminosity for NGC~4388 to be $L_{\rm bol}=8.5\times10^{43}~{\rm erg~s^{-1}}$, using our Swift BAT ${\rm 14-195~keV}$ flux density measurement in \citetalias{2021ApJ...906...88F}. This provides a kinetic luminosity of $L_{\rm wind}=8.5\times10^{40}-8.5\times10^{42}~{\rm erg~s^{-1}}$, assuming $0.1-10\%$ of the bolometric luminosity is converted into AGN winds.

To estimate the total kinetic energy from the wind, we can simply multiply the kinetic luminosity from winds by a typical lifetime for AGN activity. We followed \cite{2020ApJ...904...83S}, where they assumed a typical quasar lifetime of $10~{\rm Myr}$ during which the quasar is considered to be active. We also estimated the synchrotron lifetime from the cosmic ray electrons, as in \cite{2016ApJ...824...30D}, using
\begin{equation}
t_{\rm synch}\approx1.06\times 10^{9}~{\rm yr}\left(\frac{B}{\rm \mu G}\right)\left(\frac{\nu}{\rm GHz}\right).
\end{equation}
We again used $B_{\rm tot}=67~{\rm \mu G}$ at the nucleus for $\nu=6~{\rm GHz}$ and $\nu=10~{\rm GHz}$ and find the synchrotron cooling time to range from $t_{\rm synch}=0.6-0.8~{\rm Myr}$. Thus, we estimate the kinetic energy due to AGN winds to be $1.6\times10^{54}~{\rm erg}$ if 0.1\% of the bolometric luminosity is converted to AGN winds over $0.6~{\rm Myr}$ and $2.7\times10^{57}~{\rm erg}$ if 10\% is converted over $10~{\rm Myr}$. As the energy from mechanical feedback was calculated to be ${\rm KE}\sim10^{50}~{\rm erg}$, we find that the deposited energy indeed exceeds the minimum energy measured from the synchrotron plasma and therefore AGN wind activity producing the radio emission is plausible.

Furthermore, \cite{2014MNRAS.442..784Z} calculated an efficiency factor of $3.6\times10^{-5}$ from their 1.4 GHz observations for converting the kinetic luminosity due to winds into radio luminosity. If we assume that the flat spectrum holds below 4 GHz down to 1.4 GHz, then we can just divide our luminosity measurements by this efficiency factor to estimate the expected kinetic luminosity due to winds. When we apply this efficiency factor to each of our measurements, we find wind luminosities of $L_{\rm C,~wind}=2.1\times 10^{41}~{\rm erg~s^{-1}}$, $L_{\rm C+X,~wind}=2.4\times 10^{41}~{\rm erg~s^{-1}}$, $L_{\rm X,~wind}=2.7\times 10^{41}~{\rm erg~s^{-1}}$. This is ${\sim}0.3\%$ of the bolometric luminosity and right in line with our estimates of $L_{\rm wind}=8.5\times10^{40}-8.5\times10^{42}~{\rm erg~s^{-1}}$.

With the previous discussion in mind, we hypothesize that much of the detected radio emission at VLA resolution is due to winds interacting with the immediate densely populated neighborhood surrounding the AGN, we explore this further in the next section with respect to Component B.

\subsection{The 3D Nature of Component B}
The multi-component structure of radio emission in NGC 4388 is common in Seyfert galaxies at VLA resolution \citep[e.g.][]{1984ApJ...285..439U,1995MNRAS.276.1262K}. 
While collimated and beam-like radio structures perpendicular to the accretion disk at VLA resolution is suggestive of a radio jet propagating along the poles of the disk, as in radio-loud AGNs \citep{1995PASP..107..803U}, jet activity is not necessarily always the progenitor of the radio emission. For example, NGC~1068 is a FRAMEx target that has an elongated, multi-component radio structure at VLA resolution which has been interpreted as a jet that aligns with an ionizing radiation cone \citep{1988ApJ...328..519P,1997ApJ...476L..67C}. But recent VLBA observations by \cite{2023ApJ...953...87F} suggest that no small-scale jet is seen, and the observed radio emission is instead formed by AGN winds shocking the host environment.
They find that the nuclear radio continuum in NGC~1068 is intricately intertwined with molecular and ionized gases. As prominent [\ion{Fe}{2}] emission is indicative of shocked emission \citep{2013ApJ...777..156I}, the [\ion{Fe}{2}] emission observed in NGC~1068 is inferred as a shock distribution. This distribution is suggested to trace the surface of the molecular gas lanes interacting with the optical NLR, where shock fronts are likely formed from molecular gas lanes rotating into the AGN radiation field. Furthermore, they analyzed two VLBA observations of NGC~1068 with epochs separated by 22 years and found that the measured velocities of the radio structure is traveling at sub-relativistic speeds.

As in NGC~1068, NGC~4388 has a radio component that is also aligned with an ionization cone near the AGN in the plane of the sky. The region is complex, and Figure \ref{fig:multiwavelength} shows observations of a soft X-ray nebula that traces the forbidden [\ion{O}{3}] line. The soft X-ray feature depicted in Figure \ref{fig:xray} was studied extensively in \cite{2003MNRAS.345..369I}, where they observed a mix of emission lines between $0.5-2~{\rm keV}$.
\cite{2003MNRAS.345..369I} suspect the X-ray structure is not produced by thermal processes, but rather can be explained by a photoionized plasma. The surface brightness of the X-ray nebula softens with increasing radius and appears to be consistent with a gas expanding at a constant velocity, perhaps ejected from the AGN. Hard X-rays show that a strong circumnuclear Fe K$\alpha$ line at $6.2-6.7$ keV is linked to the AGN in NGC~4388, with extended features out to ${\sim}0.8$ kpc, and further favors the photoionization model \citep{2003MNRAS.345..369I,2021ApJ...908..156Y}. More recently, \cite{2023MNRAS.522..394Y} analyzed the NGC 4388 spectrum $>2~{\rm keV}$ using archival Suzaku observations and find that the spectrum can rather simply be explained by modeling a uniformly spherical distribution of matter. The modeling holds out to $>10^4$ gravitational radii and has Fe absorption lines indicative of a highly ionized wind, which could potentially be associated with the accretion disk.

The expanding gas scenario is supported by the alignment of the [\ion{O}{3}] emission shown Figure \ref{fig:OIII}. In the \cite{2003MNRAS.345..369I} modeling, the gas on the inner edge of the shell has a higher ionization factor, with an abundance of \ion{O}{7}, than that on the outer edge where the [\ion{O}{3}] is likely to peak. The [\ion{O}{3}] gas itself has complex structure with a region of high excitation on the western edge near the apex.

It is possible that the southwest ionization cone could mimic the multi-wavelength structure seen across the AGN {2MASX~J0243} in \cite{2019ApJ...887..200F}, which also has radio emission in a pseudo-jet morphology. There, the optical gas lanes that trace the radio and X-ray emission are ionized by a radiatively driven outflow from the AGN. While their AGN feedback models were successful at launching emission-line gases at distances near the nucleus, the models could not explain the high-velocity gases observed at larger radii. Rather, they suggested that the X-ray emission tracing the optical gas lanes is produced by high-velocity molecular gas lanes which thermally expand into X-ray winds from the AGN. The ionizing winds are subsequently shocked, producing thermalized X-ray emission and radio emission peaking at the boundaries of the impact regions. The optically ionized gases observed are thus interpreted as outflows that trace the surface of the high-velocity gas lanes, illuminated by the AGN's ionizing radiation field.

Like NGC~1068 and {2MASX~J0243}, Component B is intertwined within an ionized cavity, and we similarly suggest that the radio emission is most likely a radiative outflow that is interacting with the ISM for the following reasons. The medium and soft X-rays are in an extended distribution which appear to trace the edges of the [\ion{O}{3}] emission, as in {2MASX~J0243}. Since [\ion{O}{3}] is a forbidden line, there must be a strong radiative outflow hitting a dense region of the ISM. This is supported by the findings in \cite{2003MNRAS.345..369I}, which suggested synonymous origins for the soft X-ray nebula and [\ion{O}{3}] gas, perhaps due to an expanding gas. Expanding gas through the ISM naturally aligns with the wind-based scenario discussed in Section \ref{sec:component_A} and FRAMEx~V where the AGN winds shock the local neighborhood and in-turn produce the observed synchrotron radiation \citep{2015MNRAS.447.3612N,2019MNRAS.482.3023P}. The steep spectrum measured for Component B ($\alpha=-1.07$) is also indicative of electrons potentially losing their energy faster than the dynamical timescale of the shock.
Unlike the thermal processes observed in the X-rays at large radii discussed in \cite{2019ApJ...887..200F}, thermal collisionally-ionized radiation is not easily measured within the ionization cone due to the complex X-ray nebula. For NGC~4388, Component B is close in proximity to the AGN where photoionization processes are more likely to dominate.

An indicator of shocked processes producing the enhanced radio emission is inferred by the strong [\ion{Fe}{2}]/Pa$\beta$ flux ratio (${\gtrsim}2$) that is spatially coincident with Component B \citep{2001AJ....122..764K}. We hypothesize that if the ionization and shocking mechanisms in NGC~4388 mirror that of {2MASX~J0243} and {NGC~1068}, the ionization cone is likely projected emission on the underside of the galactic disk, and the radio emission in Component B is observed due to a shocked product. The underside of the disk showing visible shocked processes is further supported by an enhanced [\ion{Fe}{2}] emission line extending along the ionization cone to the southwest \citep{2017MNRAS.465..906R}. The [\ion{Fe}{2}] emission line is alternatively much fainter to the northeast of Component A along multiple position angles (30$\degr$ in \cite{2001AJ....122..764K}, 64$\degr$ in \cite{2017MNRAS.465..906R}) where the disk is likely obscuring the line of sight.

\section{Summary and Conclusions} \label{sec:conclusion}
We have obtained a uniform data set of a volume-complete sample of 25 AGNs out to 40 Mpc for sources with declinations between $-30\degr$ to $60\degr$ using the VLA at resolutions ${\gtrsim}0\farcs3$. Contemporaneous observations were done at $4-8$ GHz and $8-12$ GHz, providing broad $uv$-coverage and sensitivity by the VLA in its highest resolution A-array configuration and allow for morphological and spectral analysis at the sub-arcsecond level. We have uniquely obtained a uniform sample of polarized radio continuum at these resolutions to explore the magnetic field properties along the line of sight. We can use the polarized radio continuum to compare to radio jets or jet-like phenomena to provide a new facet for understanding the AGN feedback processes.

NGC~4388 is the first target analyzed for this VLA follow-up project to the VLBA observations in \cite{2021ApJ...906...88F}, and we showcase the morphological, spectral, and polarization capabilities of our calibration and imaging approach in this paper. Our conclusions are as follows:

\begin{enumerate}

\item Using the MTMFS algorithm in \textsc{casa}, we find that spectral fitting is sensitive to spatial scales and bandwidth. 
We found that the spectral index + curvature results appeared to diverge from channelized data near the edges of their respective bands for the peak flux density measurements. While wideband observations allow for resolved spectra in the radio continuum, more complex modeling or spectral fitting methods may be needed for multi-frequency synthesis with large bandwidth ratios.

\item We have detected linearly polarized radio continuum in NGC~4388 at arcsecond resolution across both Components A and B, where no strong detection has previously been made. No circular polarization was detected in the field. We find that Components A and B are weakly polarized, and the polarized radio continuum is ${<}10\%$ of the Stokes I continuum. We detect Faraday rotation for Component A (where the AGN likely resides), which is deeply embedded within the edge of the disk of the host galaxy along the line of sight. Future work will entail rotation measure synthesis to measure the Faraday dispersion and rotation of the magnetic field along the line of sight.

\item We find a broken power law for Component A across $4-12$ GHz, with a low frequency flat spectrum below a frequency break at 7.2 GHz, above which the spectrum steepens to $\alpha=-0.34$. We compared the VLA measurements to our recent VLBA results from  FRAMEx~V and find a similarly flat spectrum from $1.4-8.6$ GHz.
The VLBA peak flux densities are ${<}10\%$ of that of the VLA measurements, and we suggest that the VLA radio continuum measurements are likely due to wind interactions dominating at short radii.

\item We suggest that Component B is not a jet but rather is consistent with a radiative outflow interacting with the ISM. AGN photoionization is likely to dominate at the radius of Component B, and is evidenced by multi-wavelength studies of the ionization cone. Component B shows a steep spectral index of $\alpha=-1.07\pm0.11$ and is potentially produced via enhanced magnetic fields from wind-shocked emission, which in turn produces the observed synchrotron radiation. A strong [Fe II]/Pa$\beta$ line ratio has previously been detected to be spatially coincident with Component B, further supporting an in~situ shock-excited region. The ionization cone and Component B are suggested to be projected material on the underside of the disk of the host galaxy.

\end{enumerate}
\section*{Acknowledgements}
The authors thank the anonymous referee for their constructive feedback that greatly improved this work. We thank Theresa Wiegert, Judith Irwin and the CHANG-ES collaboration for helpful discussions and comments on their calibration and imaging methods which we used as inspiration for our work.

The National Radio Astronomy Observatory is a facility of the National Science Foundation operated under cooperative agreement by Associated Universities, Inc. The authors acknowledge use of the Very Long Baseline Array under the US Naval Observatory’s time allocation. This work supports USNO’s ongoing research into the celestial reference frame and geodesy.
\facilities{VLA, Chandra X-ray Observatory, Hubble}
\software{CASA, astropy, scipy}

\bibliography{ngc4388}{}
\bibliographystyle{aasjournal}
\appendix
\section{Stokes IQUV Maps}\label{apx:stokes}
\begin{figure*}[hbp!]
    \centering
    \includegraphics[width=\linewidth]{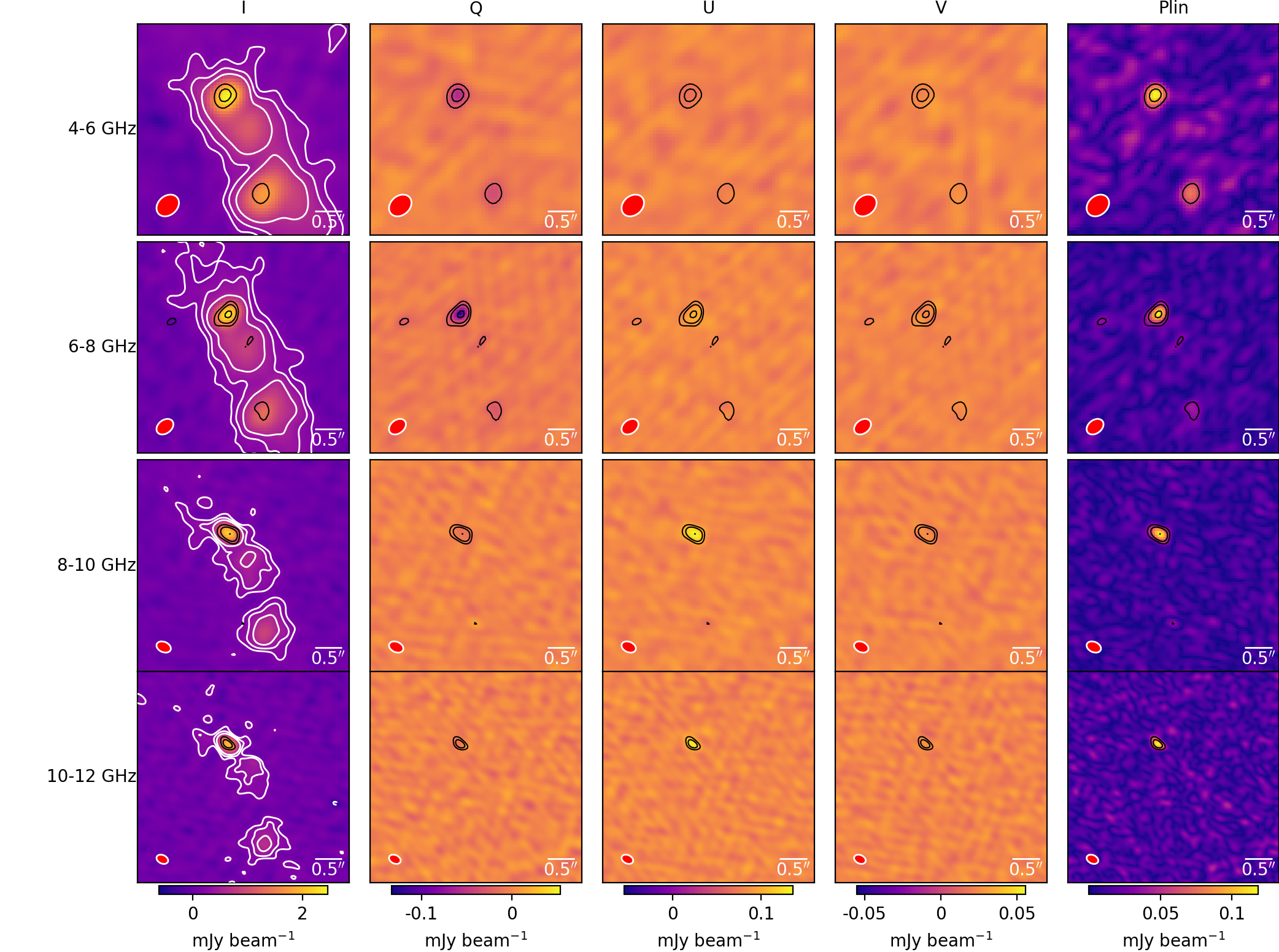}
    \caption{Stokes I, Q, U, V, and linear polarization maps for NGC 4388, for each 2~GHz baseband. The white contour lines represent $(-5,5,10,20)\times\sigma_I$ for the respective I, Q, U, and V map, while the black contour lines represent the linear polarization at the $(-5,-3,3,5,10)\times\sigma_{\rm Plin}$ levels where $\sigma_{\rm Plin}=\sqrt{\sigma^2_Q+\sigma^2_U}$.}
    \label{fig:stokes_iquv}
\end{figure*}

\section{Spectral Maps}\label{apx:spix}
\begin{figure*}[htp!]
    \begin{center}
    \subfigure[$C$-band MTMFS results for the central region of NGC~4388.]{\label{fig:spix_C}\includegraphics[width=0.8\linewidth]{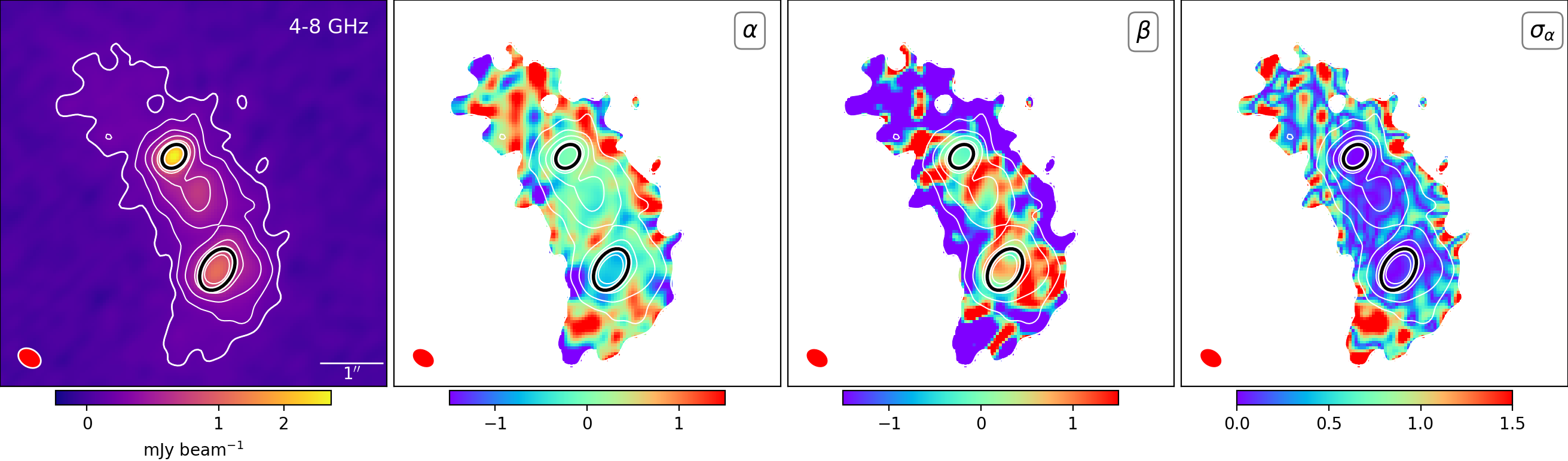}}\\
    \subfigure[$X$-band MTMFS results for the central region of NGC~4388. ]{\label{fig:spix_X}\includegraphics[width=0.8\linewidth]{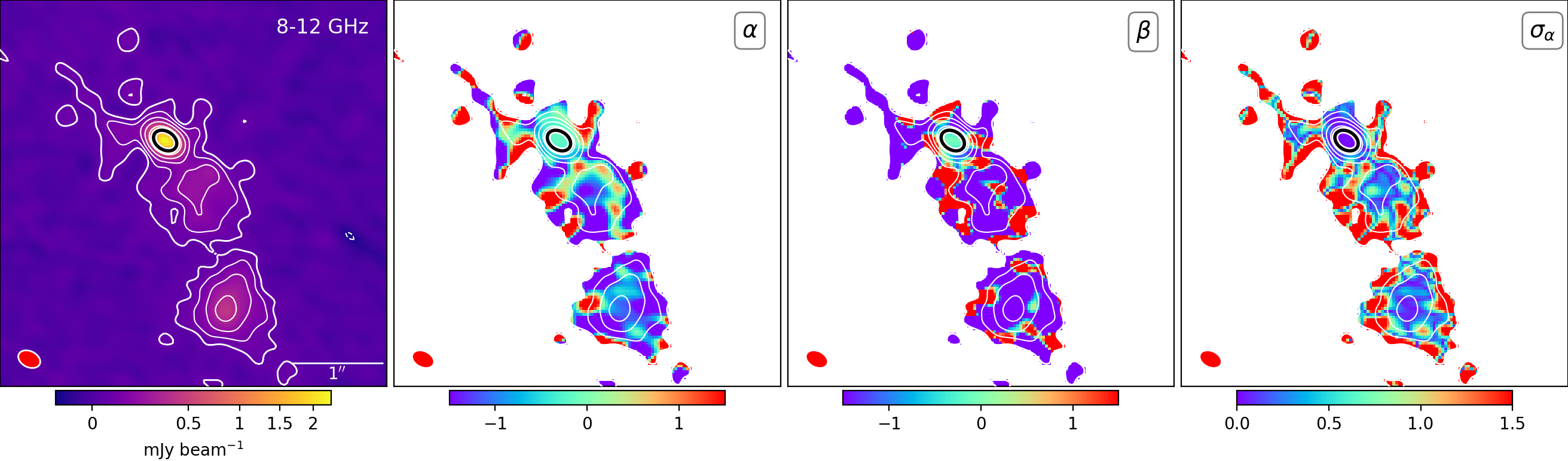}}\\
    \subfigure[$C+X$-band MTMFS results for the central region of NGC~4388 with 4 Taylor terms.]{\label{fig:spix_CX}\includegraphics[width=0.8\linewidth]{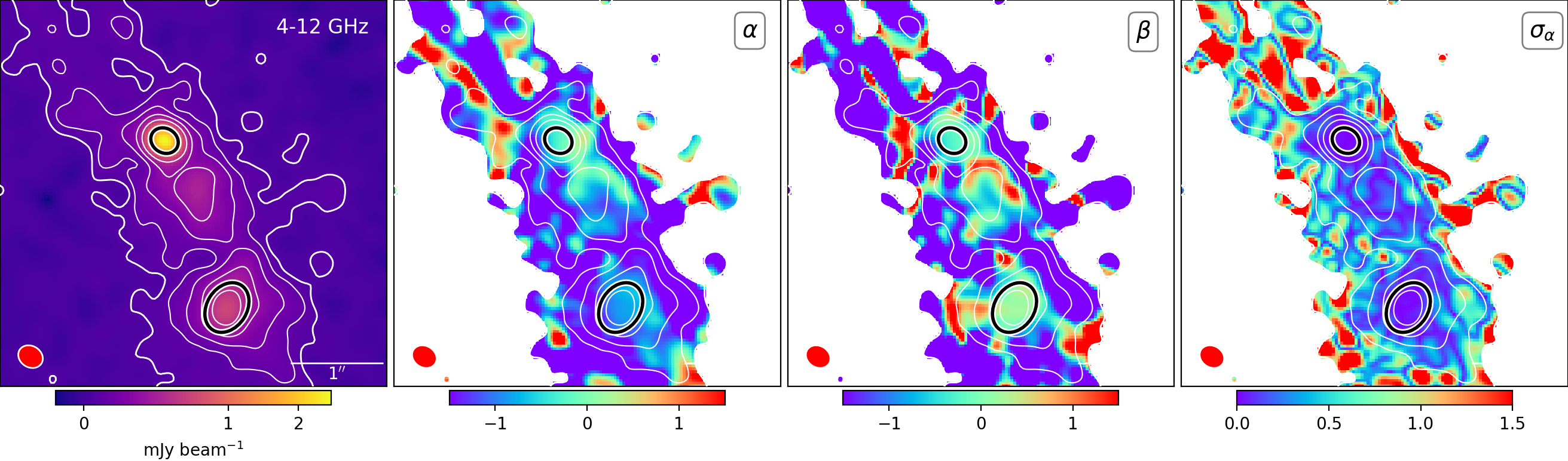}}
    \end{center}
    \caption{The white contour lines represent flux densities above $(-5,5,10,20,40,80)\times\sigma_I$. The black demarcations indicate the FWHM of the 2D Gaussian fits. The mean values within the FHWMs are indicated in Table \ref{tab:measurements}. }
    \label{fig:spix_all}
\end{figure*}

\end{document}